\documentclass[11pt,fleqn]{article}
\usepackage{epsfig}
\usepackage{rotating}
\usepackage{amsmath}
  
\usepackage{subfigure}
\renewcommand{\thesubfigure}{(\arabic{subfigure})}

\newcommand{\nnl}{\nonumber\\}
\newcommand{\kk}[1]{_{(#1)}}  
\newcommand{\kkx}[2]{_{#1(#2)}}  
\newcommand{\parity}{\mathbf{P}}
\newcommand{\mcD}{\mathcal{D}}
\newcommand{\mcE}{\mathcal{E}}
\newcommand{\mcG}{\mathcal{G}}
\newcommand{\mcL}{\mathcal{L}}
\newcommand{\mcQ}{\mathcal{Q}}
\newcommand{\mcU}{\mathcal{U}}
\newcommand{\fslash}[1]{#1\!\!\!/}  
\DeclareMathOperator{\diag}{diag} 
\newcommand{\nor}{\frac{n}{R}} 
\newcommand{\norsq}{\frac{n^2}{R^2}} 
\newcommand{\GeV}{~\text{GeV}}
\newcommand{\dpkt}{\;:\quad}
\newcommand{\sw}{s_w}  
\newcommand{\cw}{c_w}

\newcommand{\RE}{{\rm Re}}
\newcommand{\IM}{{\rm Im}}
\newcommand{\vcb}{|V_{cb}|}
\newcommand{\vtd}{|V_{td}|}
\newcommand{\vub}{|V_{ub}/V_{cb}|}
\newcommand{\vts}{|V_{ts}|}
\newcommand{\vus}{|V_{us}|}

\def\R1{\varepsilon_1}
\def\E8{\varepsilon_8}

\def\eps{\varepsilon}

\def\as{\alpha_s}

\newcommand{\mt}{m_{\rm t}}

\newcommand{\mc}{m_{\rm c}}

\newcommand{\mb}{m_{\rm b}}

\newcommand{\gev}{\, {\rm GeV}}
\newcommand{\mev}{\, {\rm MeV}}

\newcommand{\Lms}{\Lambda_{\overline{\rm MS}}}

\newcommand{\newsection}[1]{\section{#1}\setcounter{equation}{0}}
\newcommand{\bea}{\begin{eqnarray}}
\newcommand{\eea}{\end{eqnarray}}
\newcommand{\bd}{\begin{displaymath}}
\newcommand{\ed}{\end{displaymath}}

\newcommand{\be}{\begin{equation}}
\newcommand{\ee}{\end{equation}}
\newcommand{\bi}{\begin{itemize}}
\newcommand{\ei}{\end{itemize}}
\newcommand{\ord}{{\cal O}}

\newcommand{\kpnn}{K^+\rightarrow\pi^+\nu\bar\nu}
\newcommand{\kpn}{K^+\rightarrow\pi^+\nu\bar\nu}
\newcommand{\klpn}{K_{\rm L}\rightarrow\pi^0\nu\bar\nu}

\newcommand{\klm}{K_{\rm L} \to \mu^+\mu^-}
\newcommand{\kmm}{K_{\rm L} \to \mu^+ \mu^-}

\newcommand{\imlt}{\IM\lambda_t}
\newcommand{\relt}{\RE\lambda_t}
\newcommand{\relc}{\RE\lambda_c}

\renewcommand{\baselinestretch}{1.3}

\begin{document}
\thispagestyle{empty}
\phantom{xxx}
\vskip0.3truecm
\begin{flushright}
 TUM-HEP-496/02 \\
 MPI-PhT/2002-80 \\
December 2002
\end{flushright}
\vskip0.4truecm

\begin{center}
 {\LARGE\bf The Impact of Universal Extra Dimensions on\\
 the Unitarity Triangle and Rare K and B Decays\\ }
\end{center}

\vskip.6truecm
\centerline{\Large\bf Andrzej J. Buras${}^{a}$, Michael
  Spranger${}^{a,b}$ and Andreas Weiler${}^a$} 
\bigskip
\centerline{\sl ${}^a$ Physik Department, Technische Universit{\"a}t M{\"u}nchen } 
\centerline{\sl D-85748 Garching, Germany}
\centerline{\sl ${}^b$ Max-Planck-Institut f{\"u}r Physik - Werner-Heisenberg-Institut}
\centerline{\sl D-80805 M{\"u}nchen, Germany}

\vskip0.6truecm
\centerline{\bf Abstract}
We calculate the contributions of the Kaluza-Klein (KK) modes  to 
the $K_L-K_S$ mass difference $\Delta M_K$, the parameter $\varepsilon_K$, 
the $B^0_{d,s}-\bar B^0_{d,s}$ mixing mass differences $\Delta M_{d,s}$ and 
rare decays  
$\kpn$, $\klpn$, $\klm$, $B\to X_{s,d}\nu\bar\nu$ and $B_{s,d}\to\mu\bar\mu$ 
in the Appelquist, Cheng and Dobrescu (ACD) model with one universal extra 
dimension. 
For the compactification scale $1/R= 200~\gev$ the KK effects in these 
processes are governed by a $17\%$ enhancement of the $\Delta F=2$ box 
diagram function $S(x_t,1/R)$ and by a $37\%$ enhancement of the $Z^0$ 
penguin diagram function $C(x_t/1/R)$ relative to their 
 Standard Model (SM) values. This implies 
the suppressions of $\vtd$ by $8\%$, of $\bar\eta$ by $11\%$ and 
of the  angle $\gamma$  in the unitarity triangle by $10^\circ$. 
$\Delta M_s$ is increased by $17\%$. $\Delta M_K$ is essentially 
uneffected.
All branching ratios considered in this paper are increased with 
a hierarchical structure of enhancements: $\kpn~(16\%)$, $\klpn~(17\%)$, 
$B\to X_{d}\nu\bar\nu~(22\%)$, $(K_L\to\mu\bar\mu)_{\rm SD}~(38\%)$,
 $B\to X_{s}\nu\bar\nu~(44\%)$,  $B_{d}\to\mu\bar\mu~(46\%)$ and 
$B_{s}\to\mu\bar\mu~(72\%)$.
For $1/R= 250~(300)\gev$ all these
effects are decreased roughly by a  factor of $1.5~(2.0)$.
We emphasize that the GIM mechanism
assures the convergence of the sum over the KK modes in the 
case of $Z^0$ penguin diagrams and we 
give the relevant Feynman rules for the five dimensional ACD model.
We also emphasize that a consistent calculation of branching ratios has 
to take into account the modifications in the values of the CKM parameters.
As a byproduct we confirm the dominant $\ord (g_2 G_F m_t^4 R^2)$
correction from the KK modes to the $Z^0 b\bar b$ vertex calculated recently
in the large $m_t$ limit.

\newpage

\section{Introduction}
\setcounter{equation}{0}
During the last years there has been an increased interest in models with 
extra dimensions. Among them a special role play the ones with
universal extra dimensions (UED). In these models all the Standard Model 
(SM) fields are allowed to propagate in all available dimensions. Above
the compactification scale $1/R$ a given UED model becomes a higher
dimensional field theory whose equivalent description in four dimensions 
consists of the SM fields, the towers of their Kaluza-Klein (KK) partners 
and additional towers of KK modes that do not correspond to any field 
in the SM. The simplest model of this type is the Appelquist, Cheng and 
Dobrescu (ACD) model \cite{appelquist:01} with one extra universal dimension.
In this model 
the only additional free parameter relative to the SM is the compactification
scale $1/R$. Thus all the masses of the KK particles and their interactions 
among themselves and with the SM particles are described in terms of $1/R$ 
and the parameters of the SM. This economy in new parameters should be 
contrasted with supersymmetric theories and models with an extended Higgs 
sector.

A very important property of the ACD model is
the conservation of KK parity that implies the absence of tree level 
KK contributions to low energy processes taking place at scales $\mu\ll 1/R$.
In this context the  flavour changing neutral current (FCNC) processes like 
particle-antiparticle mixing and rare K and B decays are of particular 
interest. As these processes appearing in the SM first at one-loop are 
strongly suppressed,  the one-loop contributions from the KK modes to them 
could in principle be important.

The effects of the KK modes on various processes of interest have been 
investigated in a number of papers. In \cite{appelquist:01} their impact 
on the precision electroweak observables assuming 
a light Higgs ($ m_H \le 250~\gev$) led to the lower bound $1/R\ge 300~\gev$.
Subsequent analyses of the decay $B\to X_s\gamma$ \cite{AGDEWU} and of the 
anomalous magnetic moment \cite{APDO} have shown the consistency of the ACD 
model with the data for $1/R\ge 300\gev$. 
The scale of $1/R$ as low as $300\gev$ would lead to an exciting
phenomenology in the next generation of colliders 
\cite{COLL0,COLL1,COLL2,COLL3}.
Moreover the cosmic relic density 
of the lightest KK 
particle as a dark matter candidate turned out to be of the right order of
magnitude \cite{SETA}. The related experimental signatures have been 
investigated in ref. \cite{DARK}.

Very recently Appelquist and Yee \cite{APYE} have extended the analysis of 
\cite{appelquist:01} by considering  a heavy Higgs 
($ m_H \ge 250\gev$). It turns out that in this case the lower bound 
on $1/R$ can be decreased to $250\gev$, implying larger KK contributions 
to various low energy processes, in particular to the FCNC processes.
Among the latter only the decay $B\to X_s\gamma$ has been investigated 
within the ACD model so far \cite{AGDEWU} and it is desirable to consider 
other FCNC processes.

In the present paper we calculate for the first time 
the $B^0_{d,s}-\bar B^0_{d,s}$ mixing mass differences $\Delta M_{d,s}$,
the $K_L-K_S$
mass difference, the CP violation parameter $\varepsilon_K$ and the 
branching ratios for the rare decays 
$\kpn$, $\klpn$, $\klm$, $B\to X_{s,d}\nu\bar\nu$ and $B_{s,d}\to\mu\bar\mu$ 
in the ACD model with one universal extra  
dimension. In the forthcoming paper \cite{BPSW02} we will analyze 
the decays $B\to X_s\gamma$, $B\to X_s l^+l^-$ and $K_L\to \pi^0 e^+e^-$. 
In order to be more general we will include the results for 
$1/R=200\gev$ that is only slightly below the lowest value of
$1/R=250\gev$ allowed by the electroweak precision data.

As our analysis shows, the ACD model with one extra dimension has a number 
of interesting 
properties from
the point of view of FCNC processes 
discussed here. These are: 

\begin{itemize}
\item
GIM mechanism \cite{GIM} that improves significantly the convergence of 
the sum over the KK modes 
corresponding to the top quark, removing simultaneously to an excellent 
accuracy the contributions of the KK modes corresponding to lighter 
quarks and leptons. This feature removes the sensitivity of the calculated
branching ratios to the scale $M_s\gg 1/R$ at which the higher dimensional 
theory becomes non-perturbative and at which the towers of the KK particles 
must be cut off in an appropriate way. This should be contrasted with 
models with fermions localized on the brane, in which the KK parity is not 
conserved and the sum over the KK modes diverges. 
In these models the results are sensitive to $M_s$ and the KK 
effects in $\Delta M_{s,d}$ are significantly larger \cite{OLPASA} than 
found here.
\item
The low energy effective Hamiltonians are governed by local operators 
already present  in the SM. As flavour violation and CP violation in 
this model is entirely governed by the CKM matrix, the ACD model belongs 
to the class of the so-called models with minimal flavour violation (MFV) 
as defined in \cite{UUT}. This has automatically two important 
consequences.
\item
The impact of the KK modes on the processes discussed here amounts 
to the modification of the Inami-Lim one-loop functions
\cite{IL}. This is the 
function $S$ \cite{BSS} in the case of $\Delta M_{d,s}$
and of the parameter $\varepsilon_K$ and the functions $X$ and $Y$
\cite{PBE0}  in the 
case of the rare decays considered. In the ACD model these three functions 
depend only on $\mt$ and the single new parameter, the compactification 
radius $R$.
\item
The unitarity triangle constructed from $\vub$, $\Delta M_d/\Delta M_s$ 
and the $\sin 2\beta$ extracted from the CP asymmetry $a_{\psi K_S}$ is 
common to the SM model and the ACD model. That is, the $R$-dependence drops out 
in this construction. Which of these two models, if any, is consistent with 
the data can only be found out by analyzing $\Delta M_d$ and $\Delta M_s$
separately, $\varepsilon_K$ and in particular the branching ratios for rare 
K and B decays that depend explicitly on $R$.
\end{itemize}

Our paper is organized as follows. In section 2, we summarize those 
ingredients of the ACD model that are relevant for our analysis. In 
particular, we give in appendix A the set of the relevant Feynman rules
in the ACD model that have not been given so far in the literature. In 
section 3, we calculate the KK contributions to the box diagram function $S$
and we discuss the implications of these contributions for $\Delta M_K$, 
$\Delta
M_d$, $\Delta M_s$, $\varepsilon_K$ and the unitarity triangle. In
section 4, we calculate  
the corresponding corrections to the functions $X$ and $Y$ that
receive the  
dominant contribution from $Z^0$-penguins and we analyze the implications 
of these corrections for the rare decays
$\kpn$, $\klpn$, $\klm$, $B\to X_{s,d}\nu\bar\nu$ and $B_{s,d}\to\mu\bar\mu$. 
In section 5, we summarize our results and give a brief outlook.

Very recently an analysis of $\Delta M_{d,s}$ in 
the ACD model has been presented in \cite{CHHUKU}. In the first version of 
this paper the result for the 
function $S$ found by these authors differed significantly from our result
with the effect of the KK modes being by roughly a factor of two larger 
than what we find. After the first appearence of our paper 
the authors of \cite{CHHUKU} identified errors in their calculation and 
confirmed our result for $S$. However, we disagree with their 
claim that the reduction of the error on the parameter 
$\sqrt{\hat B_{B_d}}F_{B_d}$ by a factor of three will necessarily increase
the lowest allowed value of the compactification scale $1/R$ to $740\gev$. 
We will address this point at the end of section 3.

\section{The Five Dimensional ACD Model}
\setcounter{equation}{0}
The five dimensional UED model introduced by Appelquist, Cheng and
Dobrescu (ACD) in \cite{appelquist:01} uses orbifold compactification to
produce chiral fermions in 4 dimensions. This is not the case in the models 
described
in \cite{rueckl}, where all fermions are localized on the 4
dimensional brane. However, there are many similarities between
these two classes of models, and some of the issues discussed in this
section have already been presented in detail in \cite{rueckl}. We
assume vanishing boundary kinetic terms at the cut off scale. 
We also rely on \cite{appelquist:01,Georgi,Giedt}. 

\subsection{Kaluza-Klein mode expansion}
The topology of the fifth dimension is the orbifold $S^1/Z_2$, and the
coordinate $y \equiv x^5$ runs from $0$ to $2\pi R$, where $R$ is the
compactification radius. The orbifold has two fixed points, $y=0$ and
$y=\pi R$. The boundary conditions given at these fixed points
determine the Kaluza-Klein (KK) mode expansion of the fields.

A scalar field $\phi$ has to be either even or odd under the
transformation $\parity_5: y\rightarrow-y$, and therefore
\be\left.
  \begin{aligned}
    \partial_5 \phi^+ &= 0 && \quad\text{for even fields}&\\
    \phi^- &= 0            && \quad\text{for odd fields}&
  \end{aligned}
  \right\} \quad\text{at } y=0, \pi R.
\ee
These are von Neumann and Dirichlet boundary conditions respectively at
the fixed points. The associated KK expansions are
\be\label{scalarKKexpansion}
  \begin{alignedat}{2}
      \phi^+ (x,y)& = \frac{1}{\sqrt{2\pi R}} \phi^+\kk 0 (x) \,+\,&
    \frac{1}{\sqrt{\pi R}} \sum\limits_{n=1}^\infty
    \phi^+\kk n(x) \cos\frac{n y}{R},\\
    \phi^- (x,y)& =& 
    \frac{1}{\sqrt{\pi R}} \sum\limits_{n=1}^\infty
    \phi^-\kk n(x) \sin\frac{n y}{R},
  \end{alignedat}
\ee
where $x \equiv x^\mu,\;\mu = 0,1,2,3$, denotes the four non-compact
space-time coordinates. The fields $\phi^\pm\kk n (x)$ are called {
  Kaluza-Klein modes}.  

A vector field $A^M$ in 5 dimensions has five
components, $M = 0,1,2,3,5$. The orbifold compactification forces the
first four 
components to be even under $\parity_5$, while the fifth component is
odd:
\be\left.
  \begin{aligned}
    \partial_5 A^\mu &= 0& \\
    A^5 &= 0&
  \end{aligned}
  \right\} \quad\text{at } y=0, \pi R.
\ee

Hence, the KK expansion of a vector field is
\be\label{vectorKKexpansion}
  \begin{alignedat}{2}
    A^\mu(x,y) &= \frac{1}{\sqrt{2\pi R}} A^\mu\kk 0(x) \,+\,&
    \frac{1}{\sqrt{\pi R}} \sum\limits_{n=1}^\infty 
    A^\mu\kk n(x) \cos\frac{n y}{R},\\
    A^5(x,y) &= &\frac{1}{\sqrt{\pi R}} \sum\limits_{n=1}^\infty 
    A^5\kk n(x) \sin\frac{n y}{R}.
  \end{alignedat}
\ee

A Dirac spinor $\psi$ in 5 dimensions is a four component
object. Using the chirality projectors $P_{R/L} =
(1\pm\gamma_5)/2$, a spinor $\psi = (P_R + P_L) \psi = \psi_R
+ \psi_L$ has to
satisfy either
\begin{align}\label{P5}
  \left.
  \begin{aligned}
    \partial_5 \psi^+_R &= 0 & \\
    \psi^+_L &= 0 &
  \end{aligned}
  \right\}  \quad\text{at } y=0, \pi R
&&\text{or}&&
  \left.
  \begin{aligned}
    \partial_5 \psi^-_L &= 0 & \\
    \psi^-_R &= 0 &
  \end{aligned}
  \right\}  \quad\text{at } y=0, \pi R.
\end{align}

The respective KK mode expansions are
\be\label{spinorKKexpansion}
  \begin{alignedat}{2}
    \psi^+(x,y) &= \frac{1}{\sqrt{2\pi R}}\psi_{R(0)}(x) +
     \frac{1}{\sqrt{\pi R}} \sum_{n=1}^\infty \left( \psi_{R(n)}(x)
    \cos\frac{ny}{R} + \psi_{L(n)}(x)
    \sin\frac{ny}{R} \right),\\
    \psi^-(x,y) &= \frac{1}{\sqrt{2\pi R}}\psi_{L(0)}(x) +
     \frac{1}{\sqrt{\pi R}} \sum_{n=1}^\infty \left( \psi_{L(n)}(x)
    \cos\frac{ny}{R} + \psi_{R(n)}(x)
    \sin\frac{ny}{R} \right).
  \end{alignedat}
\ee

The zero-mode is either
right-handed 
or left-handed. The non-zero-modes come in chiral pairs. This
chirality structure is a natural consequence of the orbifold
boundary conditions. 

We can derive Feynman rules for the KK modes by explicitly 
integrating over the fifth dimension in the action:
\be
  S = \int d^4x \int\limits_0^{2\pi R} dy
  \,\mcL_5 = \int d^4x \,\mcL_4.
\ee

Using the KK mode expansions in (\ref{scalarKKexpansion}),
(\ref{vectorKKexpansion}) and (\ref{spinorKKexpansion}), the five dimensional
Lagrangian $\mcL_5$ reduces to the four dimensional Lagrangian
$\mcL_4$ which contains all KK modes. The field content is arranged
such that the zero-modes are the 4 dimensional SM particles, whereas
the higher modes constitute their KK excitations.  Moreover there 
are additional KK modes that do not correspond to any field in the SM.

\subsection{Universal Extra Dimensions}\label{ued}
In the UED scenarios, all fields present in the Standard Model live in
the extra dimensions, i.e. they are functions of all space-time
coordinates.

For bosonic fields, one simply replaces all derivatives and fields in
the 
SM Lagrangian by their 5 dimensional counterparts. There are
the $U(1)_Y$ gauge field $B$ and the $SU(2)_L$ gauge fields $A^a$,
as well as the $SU(3)_C$ QCD gauge fields. The Higgs doublet is
\be
  \phi = \frac{1}{\sqrt{2}} \left( \begin{array}{c} \chi^2 + i
        \chi^1\\ \psi - i \chi^3 \end{array} \right)
   = \left( \begin{array}{c} i\chi^+ \\ \frac{1}{\sqrt{2}}
       \left( \psi
         - i\chi^3 \right) \end{array} \right),
\ee
where
\be
  \chi^\pm = \frac{1}{\sqrt{2}} \left[ \chi^1 \mp i \chi^2
    \right].
\ee

It is chosen to be even under $\parity_5$ so it possesses a
zero-mode. By assigning a vacuum expectation value to that zero-mode
with the substitution $\psi \rightarrow \hat v + H$,
we can give masses to the fermions. Note that we label all parameters
of the 5 dimensional Lagrangian with a caret. It is convenient to
introduce 4 dimensional parameters that are related to their 5
dimensional counterparts by factors of $\sqrt{2\pi R}$, see
appendix~\ref{couplingconversion}. 

Fermions living in five dimensions are more involved. In order to write down a
Lorentz invariant Lagrangian, we need one 
$\Gamma$-matrix for each space-time dimension to satisfy the Clifford
algebra
\be
  \left\{ \Gamma^M, \Gamma^N \right\} = 2 g^{MN},\qquad M,N=0,1,2,3,5.
\ee

The metric $g^{MN} = \diag (1,-1,-1,-1,-1)$ is the natural extension of
the flat Minkowskian metric for one extra space dimension. For the
$\Gamma$-matrices, we take 
\be
  \begin{aligned}
    \Gamma^\mu &= \gamma^\mu,\qquad\mu = 0,1,2,3\\
    \Gamma^5   &= i \gamma_5,
  \end{aligned}
\ee
with $\gamma_5 = i \gamma^0\gamma^1\gamma^2\gamma^3$.
The 5 dimensional kinetic Lagrangian for spinor fields can now be
written as \cite{Georgi} 
\be
  \mcL_5 = i \overline{\psi} \Gamma^M\partial_M \psi =
  \overline{\psi} \left(i \fslash\partial - \gamma_5\partial_5
  \right) \psi.
\ee

Integrating over the fifth dimension yields
\be
  \mcL_4 = \int\limits_0^{2\pi R}dy ~\mcL_5 =
  \sum_{n=0}^\infty \overline{\psi}\kk n^\pm \left(i\fslash\partial
    \pm \nor \right) \psi\kk n^\pm,
\ee
where the sign of the KK mass term $n/R$ depends on the choice in
(\ref{P5}).

Note that the zero-mode remains massless
unless we apply the Higgs mechanism. Note also that all fields in the
4 dimensional  Lagrangian receive the KK mass $n/R$ on account of the
derivative operator $\partial_5$ acting on them. These tree-level
masses are shifted by radiative corrections due to gauge interactions
and boundary terms localized at the fixed points \cite{Cheng:2002iz}. Since
these corrections are a two-loop effect on the processes considered,
we will use the tree-level mass relations in our calculations.

\subsection{Gauge fixing and Goldstone mixing}
\label{gaugefixing}
In the 5 dimensional ACD model, we can use the same gauge fixing
procedure as in models in which the fermions are localized on the 4
dimensional subspace. We adopt the various gauge fixing functionals
given in \cite{rueckl} and adapt them to the case of the electroweak gauge
group $U(1)_Y\times SU(2)_L$ with one Higgs doublet:
\be
  \begin{aligned}
    \mcG_B [B^\mu, B_5, \chi^3] &= \frac{1}{\sqrt\xi} \left[
      \partial_\mu B^\mu - \xi 
      \left( \frac{\hat g'}{2} \hat v \chi^3 + \partial_5 B_5 \right)
    \right ],\\
    \mcG_A^a [A^{a\mu}, A^a_5, \chi^a] &= \frac{1}{\sqrt\xi} \left[
      \partial_\mu A^{a\mu} - 
      \xi \left( -\frac{\hat g_2}{2} \hat v \chi^a + \partial_5 A^a_5
      \right) \right],
  \end{aligned}
\ee
where $\hat g'$ and $\hat g_2$ are the respective 5 dimensional
$U(1)_Y$  
and $SU(2)_L$ gauge coupling constants. This is a natural extension of
the 4 dimensional $R_\xi$-gauge fixing.

With the gauge fixed, we can diagonalize the kinetic terms of the
bosons and finally derive expressions for the propagators.
Compared to the SM, there are the additional KK mass terms. As they
are common to all fields, their contribution to the gauge boson mass
matrix is proportional to the unity matrix. As a consequence, the
electroweak mixing angle $\theta_w$ is the same for all KK modes, and
we have
\begin{align}\label{ZAmix}
    \left( \begin{array}{c} Z^M \\ A^M \end{array} \right) &= \left(
      \begin{array}{cc} \cos \theta_w & - \sin \theta_w \\
      \sin \theta_w & \cos \theta_w \end{array} \right) \left(
    \begin{array}{c} A^{3M} \\ B^M \end{array} \right),\\[3mm]
  \label{Wmix}  
  W^{\pm M} &= \frac{1}{\sqrt{2}} \left (A^{1M} \mp i A^{2M}
    \right), 
\end{align}
where
\begin{align}
  s_w \equiv \sin \theta_w = \frac{g'}{\sqrt{g_2^2+g'^2}} &&\text{and}
  && c_w \equiv \cos \theta_w = \frac{g_2}{\sqrt{g_2^2+g'^2}}.
\end{align}

For the $\mu$-components, (\ref{ZAmix}) and (\ref{Wmix}) already
give the mass eigenstates
\be
  \begin{aligned}
    Z\kk n^\mu &= c_w A\kk n^{3\mu} - s_w B\kk n^\mu,\\
    A\kk n^\mu &= s_w A\kk n^{3\mu} + c_w B\kk n^\mu,\\ 
    W\kk n^{\pm\mu} &= \frac{1}{\sqrt{2}} \left (A\kk n^{1\mu} \mp i
      A\kk n^{2\mu} \right).
  \end{aligned}
\ee

The zero-modes $Z\kk 0^\mu$ and $W^{\pm\mu}\kk 0$ have the masses
\begin{align}
  M_Z = \frac{v}{2}\sqrt{g_2^2 + g'^2} &&\text{and} 
  && M_W = \frac{v}{2} g_2.
\end{align}

The components $Z\kkx 5n$ and $W\kkx 5n^\pm$ mix with the Higgs modes
$\chi^a\kk n$, while $A_{5(n)}$ as defined in (\ref{ZAmix})
is a mass eigenstate.

Because of the KK contribution to the mass matrix, the Higgs
components 
$\chi^\pm$ and $\chi^3$ with $n\not=0$ no longer play the role of Goldstone
bosons. Instead, they mix with $W_5^\pm$ and $Z_5$ to
form, in addition to the Goldstone modes $G^0\kk n$ and
$G^\pm\kk n$, 
three additional {\em physical} scalar modes $a^0\kk n$ and
$a^\pm\kk n$. The former allow the gauge bosons to acquire masses
without breaking gauge invariance. The propagator
terms are diagonalized by the orthogonal transformations
\begin{align}
    G\kk n^0 &= \frac{1}{M\kkx Zn} \left[ M_Z\chi\kk n^3
      -\nor Z\kkx 5n \right],
    & a\kk n^0 &= \frac{1}{M\kkx Zn} \left[ \nor \chi\kk n^3 +
      M_Z Z\kkx 5n \right],\\[1ex]
    G\kk n^\pm &= \frac{1}{M\kkx Wn} \left[ M_W\chi\kk n^\pm -
      \nor W\kkx 5n^\pm \right],
    & a\kk n^\pm &= \frac{1}{M\kkx Wn} \left[
      \nor\chi\kk n^\pm + M_W W\kkx 5n^\pm \right]
\end{align}
with $M\kkx Zn$ and $M\kkx Wn$ given in (\ref{GBmasses}).

For the zero-modes, we can identify $\chi\kk 0^3$ and
$\chi\kk 0^\pm$ as 
the Goldstone bosons that give masses to $Z^\mu\kk 0$
and $W\kk 0^{\pm\mu}$.
With increasing $n$, the contributions of $Z\kkx 5n$
and $W\kkx 5n^\pm$ dominate the Goldstone modes, while $\chi\kk n^3$
and $\chi\kk n^\pm$ provide the main fraction of $a\kk n^0$ and
$a\kk n^\pm$.
The physical Higgs $H\kk n$ does not mix with $A\kkx 5n$. The
latter constitutes an additional {\em unphysical} scalar mode which 
turns out to be the Goldstone mode for $A^\mu\kk n$ for $n \ge 1$.

Note that the fields $a^0$, $a^\pm$ and $A_5$ do not have zero
modes. Consequently, the photon $A^\mu\kk 0$ remains massless.

\subsection{Fermion-Higgs coupling}
The Yukawa coupling of the Higgs doublet to the quark fields
is a pivotal part of the Lagrangian concerning chirality.
Analogous to the SM, we write
\be\label{quarkHiggs}
  \mcL_{qH}(x,y) = \overline{\mcQ}' \hat{\lambda}_\mcD \mcD'
  \phi + \overline{\mcQ}' \hat{\lambda}_\mcU \mcU' i \sigma^2 \phi^*
  + \text{h.c.} 
\ee
with the three generation gauge eigenstates
\be\label{fullthreegen}
  \mcQ' = \left(
    \left( \begin{array}{c} \mcQ_u' \\ \mcQ_d' \end{array} \right),
    \left( \begin{array}{c} \mcQ_c' \\ \mcQ_s' \end{array} \right),
    \left( \begin{array}{c} \mcQ_t' \\ \mcQ_b' \end{array} \right)
    \right)^T,
\ee
\be
  \mcU' = \left( \begin{array}{c} \mcU_u' \\ \mcU_c' \\ \mcU_t'
    \end{array} \right),\qquad
  \mcD' = \left( \begin{array}{c} \mcD_d' \\ \mcD_s' \\ \mcD_b'
    \end{array} \right).
\ee

The $SU(2)$ doublets $\mcQ'$ are odd under $\parity_5$, while the
singlets $\mcU'$ and $\mcD'$ are even. Due to this assignment, the
zero-modes have the same chirality as the quark fields in the SM.

The fermions receive masses both through spontaneous symmetry breaking
and the KK expansion as described in section \ref{ued}. In order to
diagonalize the Yukawa couplings between fermions of equal 
charge, we apply the biunitary transformation
\begin{align}
    \mcQ_\mcU' &= S_\mcU~\mcQ_\mcU'',
    & \mcQ_\mcD' &= S_\mcD~\mcQ_\mcD'',\\
    \mcU' &= T_\mcU~\mcU'',
    & \mcD' &= T_\mcD~\mcD'',\\
\intertext{so that the resulting mass matrices}
    M_\mcU &= -\frac{v}{\sqrt{2}}~ S_\mcU^\dagger \lambda_\mcU T_\mcU, 
    & M_\mcD &= -\frac{v}{\sqrt{2}}~ S_\mcD^\dagger \lambda_\mcD T_\mcD
\end{align}
are diagonal and their eigenvalues $(m_u, m_c, m_t)$ and $(m_d, m_s,
m_b)$ are non-negative. This step is analogous to the SM and leads to
the CKM matrix $V_{\text{CKM}}=S_\mcU^\dagger S_\mcD$ which is unique
for all KK levels.
Next we apply to each flavour $f=u,c,t,d,s,b$ the unitary transformation
\be \label{UQmix}
  \left( \begin{array}{c}\mcU\kkx fn'' \\ \mcQ\kkx fn'' \end{array}\right)
  = \left( \begin{array}{cc} - \gamma_5 \cos\alpha\kkx fn &
      \sin\alpha\kkx fn \\
      \gamma_5 \sin\alpha\kkx fn & \cos\alpha\kkx fn \end{array}
  \right) 
    \left( \begin{array}{c}\mcU\kkx fn \\ \mcQ\kkx fn \end{array}
    \right)
\ee
and the same expression with $\mcU$ replaced by $\mcD$ for the
down-type quarks. The fields $\mcQ\kkx fn$, $\mcU\kkx fn$ and $\mcD\kkx fn$ 
are the mass eigenstates in 4 dimensions. The mixing angle is
\be\label{fermionmassmixingangle}
  \tan 2\alpha\kkx fn = \frac{m_f}{n/R}\qquad\text{for}\quad n\geq1.
\ee
In (\ref{UQmix}) we have used $\gamma_5$ to get both eigenvalues
positive and equal to
\be\label{fermionmassformula}
  m_{f(n)} = \sqrt{\frac{n^2}{R^2}+m_f^2}~.
\ee

In phenomenological applications we have $n/R \ge 200\GeV$ and can
therefore set all mixing angles to zero except for the top quark.

The Yukawa coupling of the leptons to the Higgs doublet is similar to
that of the quarks. The lepton doublets take the form
\be\label{fullthreegenleptons}
  \mcL' = \left(
    \left( \begin{array}{c} \mcL_{\nu_e}' \\ \mcL_e' \end{array}
    \right), 
    \left( \begin{array}{c} \mcL_{\nu_\mu}' \\ \mcL_\mu' \end{array}
    \right), 
    \left( \begin{array}{c} \mcL_{\nu_\tau}' \\ \mcL_\tau' \end{array}
    \right) 
    \right)^T,
\ee
and the singlets are
\be
  \mcE' = \left( \begin{array}{c} \mcE_e' \\ \mcE_\mu' \\ \mcE_\tau'
    \end{array} \right).
\ee

Due to their smallness, we can ignore neutrino masses in heavy flavour
physics. Hence, the neutrino singlets do not couple to the other
particles and are therefore omitted. Due to their small Yukawa masses,
we can set the mixing angles for all KK excitations of the leptons to
zero. The lepton masses are given by Eq.~(\ref{fermionmassformula}). 

The Feynman rules can be found in appendix~\ref{Feynmanrules}.

\section{\boldmath{$B^0_{d,s}-\bar B^0_{d,s}$} Mixing and 
\boldmath{$\varepsilon_K$}}
\setcounter{equation}{0}

\subsection{{\boldmath{$B^0_{d,s}-\bar B^0_{d,s}$} Mixing}}
The effective Hamiltonian for $\Delta B=2$ transitions in the SM
\cite{BJW90} can 
be generalized to the ACD model as follows
\begin{eqnarray}\label{hdb2}
{\cal H}^{\Delta B=2}_{\rm eff}&=&\frac{G^2_{\rm F}}{16\pi^2}M^2_W
 \left(V^\ast_{tb}V_{tq}\right)^2 \eta_{B}
 S(x_t,1/R)\times
\nonumber\\
& &\times \left[\alpha^{(5)}_s(\mu_b)\right]^{-6/23}\left[
  1 + \frac{\alpha^{(5)}_s(\mu_b)}{4\pi} J_5\right]  Q(\Delta B=2) + h. c.
\end{eqnarray}
Here $\mu_b=\ord(m_b)$, $J_5=1.627$,
\begin{equation}\label{qbdbd}
Q(\Delta B=2)=(\bar bq)_{V-A}(\bar bq)_{V-A}, \qquad q=d,s
\end{equation}
 with $(\bar bq)_{V-A}\equiv \bar b\gamma_\mu (1-\gamma_5)q$ and 
\cite{BJW90,UKJS}
\begin{equation}\label{etb}
\eta_B=0.55\pm0.01
\end{equation}
describes the short distance QCD corrections to which we will return below.

The function $S(x_t,1/R)$ is given as follows
\be\label{SACD}
S(x_t,1/R)=S_0(x_t)+\sum_{n=1}^\infty S_n(x_t,x_n)
\ee
where $(m_t \equiv \bar m_t(m_t)$)
\be\label{xtxn}
x_t=\frac{m_t^2}{M_W^2},\qquad x_n=\frac{m_n^2}{M_W^2},\qquad 
m_n=\frac{n}{R}
\ee 
and
\begin{equation}\label{S0}
S_0(x_t)=\frac{4x_t-11x^2_t+x^3_t}{4(1-x_t)^2}-
 \frac{3x^3_t \ln x_t}{2(1-x_t)^3}
\end{equation}
results from the usual box diagrams with $(W^\pm,t)$ and $(G^\pm,t)$ 
exchanges with the $\mt$ independent terms removed by the GIM mechanism.

\begin{figure}[hbt]
    \begin{minipage}[b]{0.5\linewidth}  
      \centering
        \includegraphics[scale=.92]{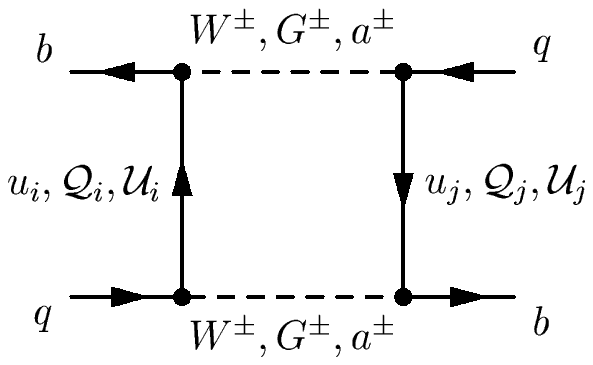} 
    \end{minipage}
    \begin{minipage}[b]{0.5\linewidth}
      \centering
        \includegraphics[scale=.92]{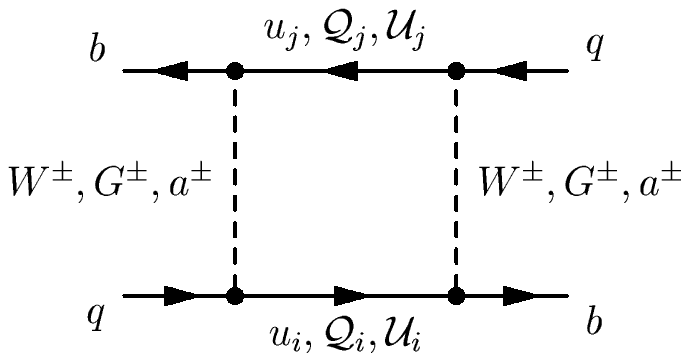} 
    \end{minipage}
    \vspace{0.1cm}
    \caption[]{\small \label{bb_box} Box diagrams contributing to $S_n(x_t,x_n)$.
     We suppress the KK mode number.} 
  \end{figure}

\begin{figure}[hbt]\renewcommand{\thesubfigure}{\space(\alph{subfigure})} 
\centering
  \subfigure[]{\label{Snplot}
      \includegraphics[scale=0.95]{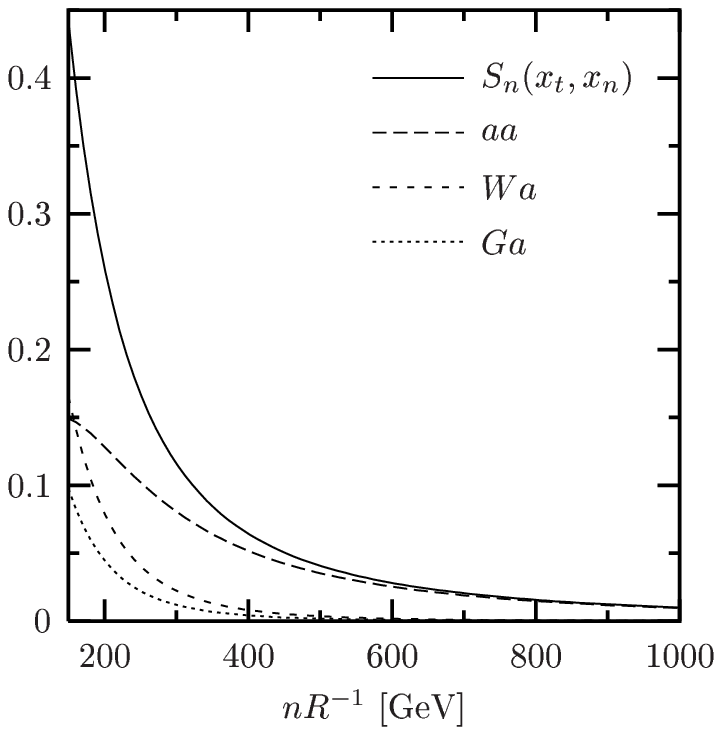} }\hspace{0.4cm}
   \subfigure[]{\label{SS0plot}
        \includegraphics[scale=.95]{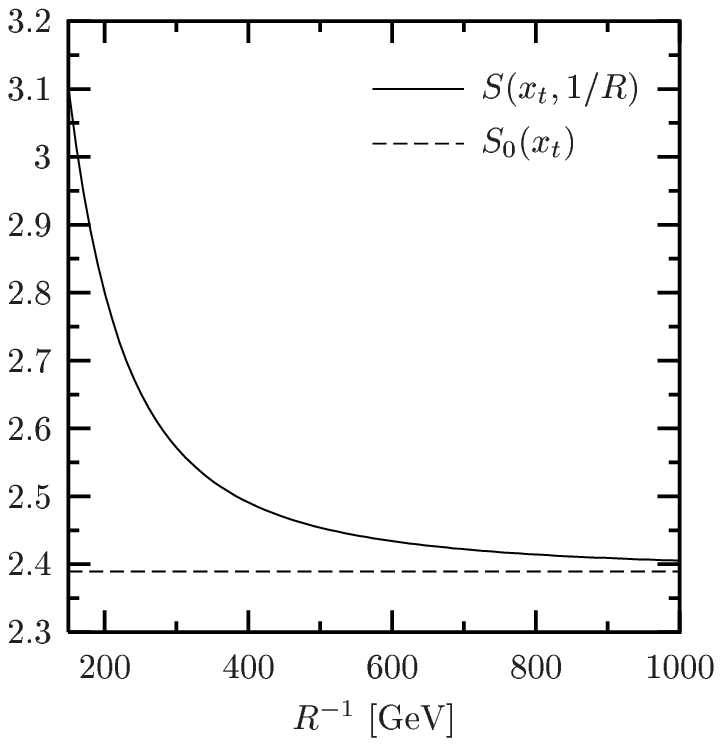} }
    \caption[]{\small (a) Contribution $S_n$ of the $n^{\rm th}$ KK mode to
  $S(x_t,1/R)$. The contributions with $a^\pm$ dominate, those with
  only $G^\pm$ and $W^\pm$ are negligible and not shown. (b) The functions
  $S(x_t,1/R)$ and $S_0(x_t)$. }
  \end{figure}

The KK contributions are represented by the functions $S_n(x_t,x_n)$ 
that are obtained by calculating the box diagrams in fig.~\ref{bb_box} with 
$W^\pm_{(n)}$, $a_{(n)}^\pm$, $G_{(n)}^\pm$, 
$\mcQ\kkx in$ and $\mcU\kkx in$ ($i=u,c,t$)
exchanges and  multiplying the result by $i/4$, where $1/4$ is a 
combinatorial factor. 
We neglect momenta and masses of external quarks.
Denoting the contribution of the sum of the 
diagrams corresponding to a given 
pair $(m_{i(n)},m_{j(n)})$ to $S_n(x_t,x_n)$ by $F(x_{i(n)},x_{j(n)})$ with 

\be\label{xin}
x_{i(n)}=\frac{m^2_{i(n)}}{M_{W(n)}^2}
\ee
and $m_{i(n)}$ and $M_{W(n)}$ defined in (\ref{fermionmassformula}) and 
(\ref{GBmasses}), and using the unitarity of the 
CKM matrix, we have
\be\label{SN}
S_n(x_t,x_n)\equiv F(x_{t(n)},x_{t(n)})+F(x_{u(n)},x_{u(n)})
-2 F(x_{t(n)},x_{u(n)})~.
\ee
As with increasing $n$ the modes $\mcQ\kkx tn$, $\mcU\kkx tn$,
$\mcQ\kkx un$ and $\mcU\kkx un$ become increasingly 
degenerate in mass and 
\be
x_{t(n)}\to x_{u(n)}\to 1,
\ee
the functions $S_n(x_t,x_n)$ decrease with increasing $n$
so that only a few terms in the sum in 
(\ref{SACD}) are relevant.  
This is seen in fig.~\ref{Snplot} where 
we show $S_n(x_t,x_n)$ as a function of $n/R$.
We will discuss this in more detail in section~\ref{GIMandConvergence}.

The contributions from different sets of diagrams to the functions 
$F(x_{i(n)},x_{j(n)})$ are collected in  appendix B. It turns out 
that the contribution from the pair $(a_{(n)}^\pm,a_{(n)}^\pm)$ is by far 
dominant. We illustrate this in fig.~\ref{Snplot}. 
In phenomenological applications 
it is more useful to work with the variables $x_t$ and $x_n$ than with 
$x_{i(n)}$. We find 
\begin{equation}\label{SFIN}\begin{split}
S_n(x_t,x_n)&=\frac{1}{4 { ( x_t-1  ) }^3 x_t}
   \bigg[6 x_n x_t - 5 x_t^2 - 
  12 x_n x_t^2 + 15 x_t^3 + 
  10 x_n x_t^3 - 11 x_t^4 - 
  4 x_n x_t^4 + x_t^5\\&\quad
  -2 x_n ( x_t -1  )^3 
   ( 3 x_n + 3 x_n x_t -x_t 
      ) 
  \ln \frac{x_n}{1 + x_n} +
   \Big(-6 x_n^2 + 2 x_n x_t + 
  12 x_n^2 x_t \\&\quad - 
  6 x_n x_t^2 - 2 x_t^3 + 
  14 x_n x_t^3 - 
  2 x_n^2 x_t^3 + 6 x_t^4 - 
  2 x_n x_t^4 \Big)
  \ln \frac{x_t + x_n}{1 + x_n}
   \bigg]~.
\end{split}
\end{equation}
In fig. \ref{SS0plot} we 
plot $S(x_t,1/R)$  versus $1/R$. For $1/R=200~\gev$ we observe 
a $17\%$ enhancement of the function $S$ with respect to its SM value given 
by $S_0(x_t)$. 
For $1/R=250\gev$ this enhancement decreases to $11\%$ and it is only 
$4\%$ for $1/R=400\gev$.

Proceeding as in the SM, we can calculate the mass differences
$\Delta M_q$ by means of
\begin{equation}\label{DMQ}
\Delta M_q = \frac{G_{\rm F}^2}{6 \pi^2} \eta_B m_{B_q} 
(\hat B_{B_q} F_{B_q}^2 ) M_W^2 S(x_t,1/R) |V_{tq}|^2,
\end{equation}
where $F_{B_q}$ is the $B_q$-meson decay constant and
$\hat B_q$ the renormalization group invariant parameter related 
to the hadronic matrix element of the operator $Q(\Delta B=2)$, see 
\cite{Erice} for details.
This implies two useful formulae 
\begin{equation}\label{DMD}
\Delta M_d=
0.50/{\rm ps}\cdot\left[ 
\frac{\sqrt{\hat B_{B_d}}F_{B_d}}{230\mev}\right]^2
\left[\frac{\vtd}{7.8\cdot10^{-3}} \right]^2 
\left[\frac{\eta_B}{0.55}\right]  
\left[\frac{S(x_t,1/R)}{2.34}\right] 
\end{equation}
and
\begin{equation}\label{DMS}
\Delta M_{s}=
18.4/{\rm ps}\cdot\left[ 
\frac{\sqrt{\hat B_{B_s}}F_{B_s}}{270\mev}\right]^2
\left[\frac{\vts}{0.040} \right]^2
\left[\frac{\eta_B}{0.55}\right]
\left[\frac{S(x_t,1/R)}{2.34}\right]~.
\end{equation}
The implications of these results for $\vtd$, $\Delta M_s$ and the
unitarity triangle will be discussed below. 

Finally, a few comments regarding the QCD factor $\eta_B$ are in order. 
This factor as given in (\ref{etb}) has been calculated within the SM 
including NLO QCD corrections that are mandatory for the proper matching
of the Wilson coefficient of the operator $Q(\Delta B=2)$ with its hadronic 
matrix element represented by the parameter $\hat B_{B_{s,d}}$ and calculated 
by means of non-perturbative methods. As the top quark and the KK modes are 
integrated out at a single scale $\mu_t=\ord(\mt,1/R)$, the contributions 
to $\eta_B$ from scales lower than $\mu_t$ are the same for the SM and the 
ACD model. They simply describe the finite renormalization of $Q(\Delta B=2)$
from scales $\ord(\mu_t)$ down to the scales $\ord(m_b)$. The difference 
between QCD corrections to the SM contributions and to the KK contributions 
arises only in the full theory  at scales $\mu_t=\ord(\mt,1/R)$ as the unknown 
QCD corrections to the box diagrams in fig. \ref{bb_box} can in principle 
differ from 
the known QCD corrections to the SM box diagrams \cite{BJW90,UKJS} that have 
been included in
$\eta_B$. As the QCD coupling constant $\alpha_s(\mu_t)$ is small and the 
QCD corrections to the SM box diagrams of order of a few percent, we do not 
expect that the difference between the QCD corrections to the diagrams in 
fig.~\ref{bb_box}  and to the SM box diagrams
is relevant, in particular in view of the fact that the 
KK contributions amount to at most $17\%$ of the full result.

For $\mu\gg \mu_t$, $\alpha_s(\mu)$ in the ACD model becomes larger 
\cite{DIDUGH} and the 
QCD corrections to KK modes with $n\gg 1$ could in principle be substantial. 
However, as seen in fig. \ref{Snplot}, these heavy modes give only a tiny 
contribution to $S(x_t,1/R)$ and can be safely neglected.

\subsection{\boldmath{$\varepsilon_K$}}
The effective Hamiltonian for $\Delta S=2$ transitions is given in the 
ACD model as follows
\begin{eqnarray}\label{hds2}
{\cal H}^{\Delta S=2}_{\rm eff}&=&\frac{G^2_{\rm F}}{16\pi^2}M^2_W
 \left[\lambda^2_c\eta_1 S_0(x_c)+\lambda^2_t \eta_2 S(x_t,1/R)+
 2\lambda_c\lambda_t \eta_3 S_0(x_c, x_t)\right] \times
\nonumber\\
& & \times \left[\as^{(3)}(\mu)\right]^{-2/9}\left[
  1 + \frac{\as^{(3)}(\mu)}{4\pi} J_3\right]  Q(\Delta S=2) + h. c.
\end{eqnarray}
where
$\lambda_i = V_{is}^* V_{id}^{}$,
  $\as^{(3)}$ is the strong coupling constant
in an effective three flavour theory and $J_3=1.895$ in the NDR scheme 
 \cite{BJW90}.
In (\ref{hds2}),
the relevant operator
\begin{equation}\label{qsdsd}
Q(\Delta S=2)=(\bar sd)_{V-A}(\bar sd)_{V-A},
\end{equation}
is multiplied by the corresponding Wilson coefficient function.
This function is decomposed into a
charm-, a top- and a mixed charm-top contribution.
The SM function $S_0(x_c,x_t)$ is defined by
\begin{equation}\label{SXCXT}
S_0(x_c,x_t)=F(x_c,x_t) + F(x_u,x_u)
- F(x_c,x_u) - F(x_t,x_u),
\end{equation}
where $F(x_i,x_j)$ is the true function
  corresponding to the box diagrams with $(i,j)$ exchanges.
One has
\begin{equation}\label{BFF}
S_0(x_c, x_t)=x_c\left[\ln\frac{x_t}{x_c}-\frac{3x_t}{4(1-x_t)}-
 \frac{3 x^2_t\ln x_t}{4(1-x_t)^2}\right]\,,
\end{equation}
where we keep only linear terms in $x_c\ll 1$, but of 
course all orders in $x_t$.

In view of the comments made after (\ref{SN}) and the structure of 
(\ref{SXCXT}), the impact of the KK modes 
on the charm- and mixed charm-top contributions is totally negligible 
and we take into account these modes only in the top contribution that 
is described 
by the same function $S(x_t,1/R)$ as in the case of $\Delta M_q$.
 This also means that the $K_L-K_S$ mass difference, $\Delta M_K$, 
being dominated 
by internal charm contributions in (\ref{hds2}) is practically uneffected 
by the KK modes.

Short distance QCD effects are described through the correction
factors $\eta_1$, $\eta_2$, $\eta_3$ and the explicitly
$\alpha_s$-dependent terms in (\ref{hds2}). 
The NLO values of $\eta_i$ are given as follows \cite{HNa,BJW90,HNb}:
\begin{equation}
\eta_1=1.45\pm 0.38,\qquad
\eta_2=0.57\pm 0.01,\qquad
  \eta_3=0.47\pm0.04~.
\end{equation}

The standard procedure allows now to calculate the CP-violating parameter 
$\varepsilon_K$ \cite{Erice}
\begin{equation}
\eps_K=C_{\eps} \hat B_K \IM\lambda_t \left\{
\RE\lambda_c \left[ \eta_1 S_0(x_c) - \eta_3 S_0(x_c, x_t) \right] -
\RE\lambda_t \eta_2 S(x_t,1/R) \right\} \exp(i \pi/4)\,,
\label{eq:epsformula}
\end{equation}
where  
$C_\eps=3.837 \cdot 10^4 $ is a numerical constant.
$\hat B_K$ is the renormalization group invariant parameter related 
to the hadronic matrix element of the operator $Q(\Delta S=2)$, see 
\cite{Erice} for details.

\subsection{Unitarity Triangle in the ACD Model}
\begin{figure}[hbt]

\centering
\includegraphics[scale=.42]{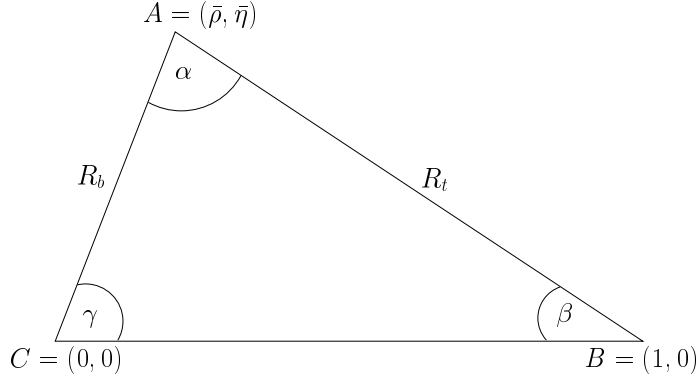} 
\caption[]{\small Unitarity Triangle.}\label{fig:utriangle}
\end{figure}

What is the impact of the KK contributions to the function $S$ on the 
elements of the CKM matrix and in particular on the shape of the unitarity 
triangle? 
In order to answer this question let us recall a few aspects of the unitarity
triangle (UT) shown in fig.~\ref{fig:utriangle} and of the Wolfenstein 
parametrization \cite{WO} as generalized to higher orders in $\lambda$ 
in \cite{BLO}. The apex of the unitarity triangle is given by \cite{BLO}
\begin{equation}\label{2.88d}
\bar\varrho=\varrho (1-\frac{\lambda^2}{2}),
\qquad
\bar\eta=\eta (1-\frac{\lambda^2}{2}).
\end{equation}
Here $\lambda$, A, $\varrho$ and $\eta$ are the Wolfenstein 
parameters \cite{WO}.
Moreover, one has 
\be\label{VUS}
V_{us}=\lambda+\ord(\lambda^7),\qquad 
V_{ub}=A \lambda^3 (\varrho-i \eta), \qquad 
V_{cb}=A\lambda^2+\ord(\lambda^8),
\ee
\begin{equation}\label{2.83d}
 V_{ts}= -A\lambda^2+\frac{1}{2}A\lambda^4[1-2 (\varrho+i\eta)],
\qquad V_{td}=A\lambda^3(1-\bar\varrho-i\bar\eta)~.
\end{equation}

The lengths $R_b$ and $R_t$ are given by
\begin{equation}\label{2.94}
R_b \equiv \frac{| V_{ud}^{}V^*_{ub}|}{| V_{cd}^{}V^*_{cb}|}
= \sqrt{\bar\varrho^2 +\bar\eta^2}
= (1-\frac{\lambda^2}{2})\frac{1}{\lambda}
\left| \frac{V_{ub}}{V_{cb}} \right|,
\end{equation}
\begin{equation}\label{2.95}
R_t \equiv \frac{| V_{td}^{}V^*_{tb}|}{| V_{cd}^{}V^*_{cb}|} =
 \sqrt{(1-\bar\varrho)^2 +\bar\eta^2}
=\frac{1}{\lambda} \left| \frac{V_{td}}{V_{cb}} \right|
\end{equation}
and the angles $\beta$ and $\gamma$ of the UT are related
directly to the complex phases of the CKM elements $V_{td}$ and
$V_{ub}$, respectively, through
\be\label{e417}
V_{td}=|V_{td}|e^{-i\beta},\quad V_{ub}=|V_{ub}|e^{-i\gamma}.
\ee
The five constraints on the UT that we have at our disposal at 
present are:
\begin{itemize}
\item
The $R_b$ Constraint: As seen in (\ref{2.94}), the length of the side AC 
is determined from $\vub$. 
\item 
$\varepsilon_K$--Hyperbola (Indirect CP Violation in $K_L\to \pi\pi$) 
obtained from (\ref{eq:epsformula}) and the experimental value for 
$\varepsilon_K$ ($\lambda=0.221$):
\begin{equation}\label{100a}
\bar\eta \left[(1-\bar\varrho) A^2 \eta_2 S(x_t,1/R)
+ P_c(\varepsilon) \right] A^2 \hat B_K = 0.214~,
\end{equation}
where  
$P_c(\varepsilon)=0.28\pm0.05$ \cite{HNa,HNb} 
represents the charm contribution that is not 
affected by the KK contributions.
\item
$B^0_d-\bar B^0_d$--Mixing Constraint ($\Delta M_d$): 
\begin{equation}\label{106}
 R_t= \frac{1}{\lambda}\frac{|V_{td}|}{\vcb} = 0.86 \cdot
\left[\frac{|V_{td}|}{7.8\cdot 10^{-3}} \right] 
\left[ \frac{0.041}{\vcb} \right]
\end{equation}
with
\begin{equation}\label{VT}
\vtd=
7.8\cdot 10^{-3}\left[ 
\frac{230\mev}{\sqrt{\hat B_{B_d}}F_{B_d}}\right]
\sqrt{\frac{\Delta M_d}{0.50/{\rm ps}}} 
\sqrt{\frac{0.55}{\eta_B}}\sqrt{\frac{2.34}{S(x_t,1/R)}}
\end{equation}
where $\Delta M_d=(0.503\pm0.006)/{\rm ps}$ \cite{ref:lepbosc}. 
\item 
$B^0_s-\bar B^0_s$--Mixing Constraint ($\Delta M_d/\Delta M_s$): 
\be\label{Rt}
R_t=0.87~\sqrt{\frac{\Delta M_d}{0.50/{\rm ps}}}
\sqrt{\frac{18.4/{\rm ps}}{\Delta M_s}}\left[\frac{\xi}{1.18}\right],
\qquad \xi=\frac{\sqrt{\hat B_s}F_{B_s}}{\sqrt{\hat B_d}F_{B_d}}
\ee
where $\Delta M_s>14.4/{\rm ps}$ at $95\%$ C.L. \cite{ref:lepbosc}. 
\item
The direct measurement of $\sin 2\beta$ through the CP asymmetry 
$a(\psi K_S)$ in $B_d\to\psi K_S$ that is not affected by the KK
contributions. 
\end{itemize}
The main uncertainties in this analysis originate in the theoretical 
uncertainties in the non-perturbative parameters $\hat B_K$ and 
$\sqrt{\hat B_d}F_{B_d}$ and to a lesser extent in $\xi$ \cite{lellouch}: 
\be\label{lat}
\hat B_K=0.86\pm0.15, \qquad  \sqrt{\hat B_d}F_{B_d}=(235^{+33}_{-41})~
{\rm MeV},
\qquad \xi=1.18^{+0.13}_{-0.04}.
\ee
Also the uncertainties in $\vub$ and 
$\sqrt{\hat B_s}F_{B_s}=(276\pm38)~{\rm MeV}$ are substantial.
The QCD sum rules results for 
the parameters in question are similar and can be found in \cite{Jamin}.

With these formulae at hand let us enumerate a few 
general properties of the values of the CKM elements within the ACD model.
These are:
\begin{itemize}
\item
$\vus$, $\vcb$ and $|V_{ub}|$ are usually determined from tree level decays. 
As in the ACD model there are no KK contributions at the tree level, the 
absolute values of these three CKM elements are to an excellent approximation 
the same as in the SM model. From the point of view of the unitarity triangle
(UT), this means that the lengths of its two sides, 
AC and CB are common to the SM model and the ACD model.
In our numerical analysis we will use, as in \cite{BUPAST},
\begin{equation}\label{vcb}
|V_{us}| = \lambda =  0.221 \pm 0.002\,
\quad\quad
\vcb=(40.6\pm0.8)\cdot 10^{-3},
\end{equation}
\begin{equation}\label{v13}
\frac{|V_{ub}|}{\vcb}=0.089\pm0.008, \quad\quad
|V_{ub}|=(3.63\pm0.32)\cdot 10^{-3}.
\end{equation}
\item
The angle $\beta$ of the UT has been determined  recently by the 
BaBar \cite{BaBar}
and Belle \cite{Belle} collaborations 
from the CP
asymmetry $a_{\psi K_S}$ in $B\to \psi K_S$ with a high accuracy giving the
world grand average \cite{NIR02}
\be
(\sin 2\beta)_{\psi K_S}=0.734\pm 0.054~.
\label{ga}
\ee
As there 
are no new complex phases in the ACD model beyond the KM phase, the angle 
$\beta$ as extracted by means of $a_{\psi K_S}$ is common to both models in
question.
\item
A similar comment applies to $\vtd$ or equivalently the length $R_t$ of the 
side AB in fig. \ref{fig:utriangle}, when $R_t$ is extracted from the ratio 
$\Delta M_d/\Delta M_s$ that is independent of $S(x_t,1/R)$ as seen in   
(\ref{Rt}).
\end{itemize}
Thus when $\vub$, $\sin 2\beta$ from $a_{\psi K_S}$ and 
$\Delta M_d/\Delta M_s$ are used to construct the UT, there is no difference 
between the SM and the ACD model as all explicit dependence on $1/R$ 
cancels out. This universal UT (UUT) \cite{UUT} that is valid for all MVF 
models, as defined in \cite{UUT}, has recently been determined 
\cite{BUPAST,DAGIISST}.

Now, even though there exists a UUT common to the SM and the ACD model, in view
of the fact that $S(x_t,1/R)\not=S_0(x_t)$, only one of these two models, if 
any, will have $\varepsilon_K$, $\Delta M_d$ and $\Delta M_s$ that agree with
the experimental data. Let us consider $\Delta M_s$ first. As 
seen in (\ref{2.83d}) 
$\vts$ is very 
close to $\vcb$ because of  CKM unitarity. Therefore it is common with an 
excellent accuracy to both models and consequently
\be
\frac{(\Delta M_s)_{\rm ACD}}{(\Delta M_s)_{\rm SM}}=
\frac{S(x_t,1/R)}{S_0(x_t)}>1 .
\ee
However, this ratio is at most $1.17$ and the distinction between these 
two models will only be possible provided $\hat B_{B_s} F_{B_s}^2$ can 
be calculated to better than $10\%$ accuracy. A very difficult task.

\begin{figure}[]
\renewcommand{\thesubfigure}{\space(\alph{subfigure})} 
\centering  
   \subfigure[]{\label{Vtdplot}
        \includegraphics[scale=.90]{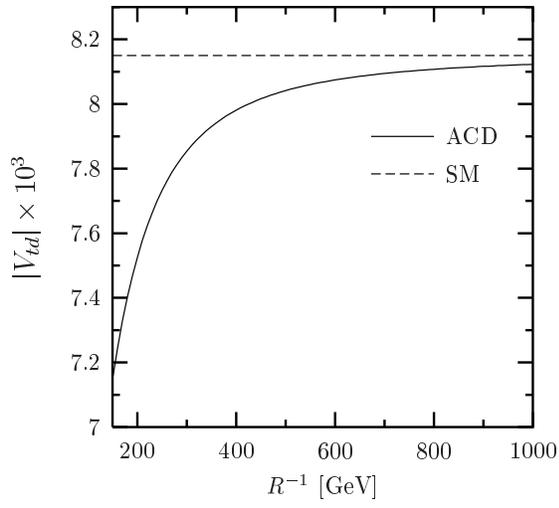} }\hspace{0.4cm}
   \subfigure[]{\label{gammaplot}
        \includegraphics[scale=.90]{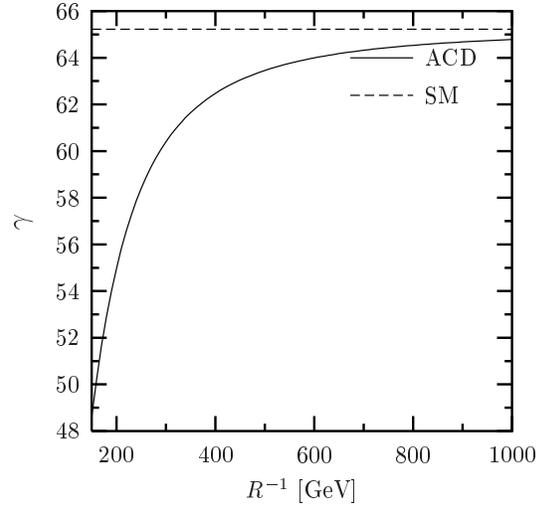} }
   \subfigure[]{\label{etabarplot}
        \includegraphics[scale=.90]{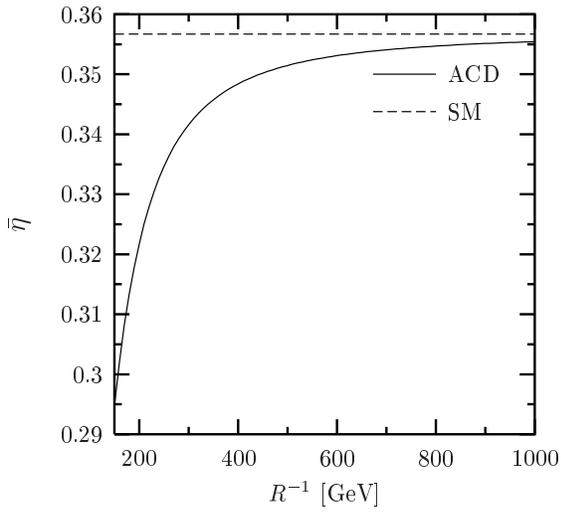} }\hspace{0.4cm}
   \subfigure[]{\label{rhobarplot}
        \includegraphics[scale=.90]{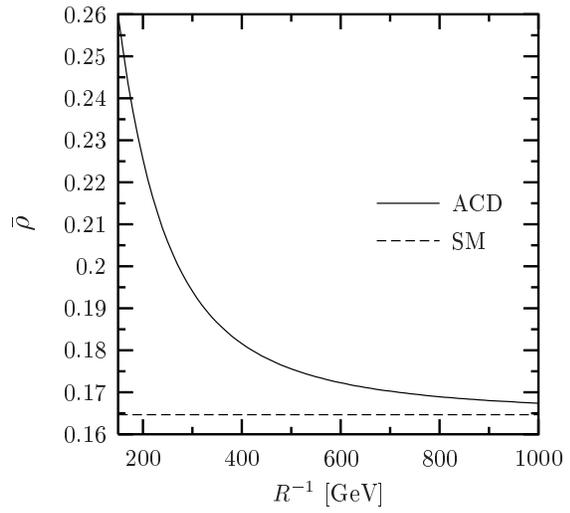} }
    \caption[]{\small\label{ckmplots} Results for various CKM
  parameters in the ACD model and in the SM: (a) $\left|V_{td}\right|$
  (b) $\gamma$ (c) $\bar\eta$ (d) $\bar \rho$.}
  \end{figure}

The fact that  $S(x_t,1/R)>S_0(x_t)$ implies also that with the
experimentally known values of $\varepsilon_K$ and $\Delta M_d$, one
has
\be
\vtd_{\rm ACD}< \vtd_{\rm SM}, \qquad  \gamma_{\rm ACD}<\gamma_{\rm SM}
\ee
as can easily be inferred from (\ref{100a}) and (\ref{VT}). In 
particular (\ref{VT}) implies 
\be\label{VTDACD}
\frac{\vtd_{\rm ACD}}{\vtd_{\rm SM}}=\sqrt{\frac{S_0(x_t)}{S(x_t,1/R)}}.
\ee
Thus $\vtd_{\rm ACD}$ can be smaller than $\vtd_{\rm SM}$ by at most 
$8\%$. In order to determine such a difference, more accurate information 
on the unitarity triangle and the non-perturbative parameters entering 
$\varepsilon_K$ and $\Delta M_{d,s}$ is necessary. Similarly 
$\gamma_{\rm ACD}$ can be smaller than $\gamma_{\rm SM}$ by at most 
$10$ degrees.

\begin{figure}[hbt]
\centering
\includegraphics[scale=0.9]{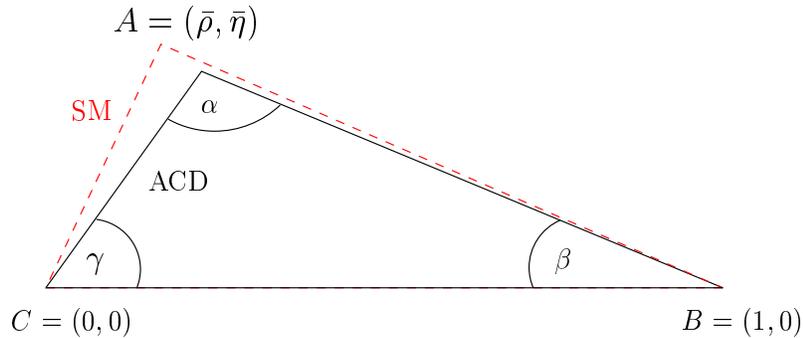} 
\caption[]{\small\label{fig:ACDutriangle} 
Unitarity Triangle in the ACD model for $1/R=200\gev$ and in the SM.}
\end{figure}

We illustrate these properties in fig.~\ref{ckmplots}, where we show 
the $1/R$ dependence of $\vtd$, $\gamma$, $\bar\varrho$ and $\bar\eta$. 
In obtaining these results we used the following procedure. First using 
the central value $\vtd_{\rm SM}=0.00815$ from the SM fit of \cite{BUPAST}
and $m_t=167\gev$ we 
determined $\vtd_{\rm ACD}$ by means of (\ref{VTDACD}). The result is
shown in fig.~\ref{Vtdplot}. Using next the central values in 
(\ref{vcb}) and (\ref{v13}) and the $1/R$ dependence of 
$(R_t)_{\rm ACD}=\vtd_{\rm ACD}/(\lambda\vcb)$, we determined  
$\gamma$, $\bar\varrho$ and $\bar\eta$ as functions of $1/R$.
We show in fig.~\ref{fig:ACDutriangle} the unitarity triangles corresponding 
to the 
ACD model with $1/R=200\gev$ and the SM model. 

\vspace{0.5cm}

In summary, the CKM elements in the ACD model extracted from  $\Delta F=2$ 
transitions and the CP asymmetry $a_{\psi K_S}$ are not expected to differ 
substantially from the corresponding values  found within the SM. This is 
very fortunate as the most recent fits of the UT 
\cite{BUPAST,NIR02,C00,achille,AJB02,Lacker}
 based on the SM expressions 
for $\Delta F=2$ transitions agree very well with the direct measurement of 
the angle $\beta$ by means of $a_{\psi K_S}$. With improved data on 
$a_{\psi K_S}$ and in particular $\Delta M_s$ and improved values for 
$\hat B_K$ and $\hat B_{B_q} F_{B_q}^2 $, a constraint on the 
compactification radius $R$ from 
$\Delta F=2$ processes is in principle possible. However, for the time 
being even the lowest value $1/R=200~\gev$ considered by us is consistent 
with the present fits of the UT. 
Setting  $1/R=200~\gev$ and repeating the analysis of \cite{BUPAST}, 
that uses the bayesian method \cite{C00},
one finds \cite{PAST} 
the values for $(\bar\varrho,\bar\eta)$, $\sin 2\alpha$, $\gamma$, 
$\Delta M_s$ and $\vtd$ in the second column of table~\ref{mfv}. 
For a comparison we give the corresponding ranges in the SM.
To this end all 
input parameters of \cite{BUPAST} have been used.

Comparing the two columns in table~\ref{mfv}, we observe all the patterns 
shown in fig.~\ref{ckmplots}. However, due to substantial uncertainties 
in the input parameters,  the effects of the  KK modes are partly washed 
out. In particular, the suppressions of $\vtd$ and $\bar\eta$ amount 
to $4-5\%$ rather than $\ord(10\%)$ that we found in fig.~\ref{ckmplots}.
This is easy to understand. With no uncertainty in 
$\sqrt{\hat B_{B_d}}F_{B_d}$,
the value of $\vtd$ is strongly correlated with $S(x_t,1/R)$ by means 
of (\ref{DMD}) as $\Delta M_d$ is known very well experimentally. 
Once the uncertainties in $\sqrt{\hat B_{B_d}}F_{B_d}$ are taken into 
account, the enhancement of the function $S$ in the ACD model can be 
compensated by the decrease of both $\vtd$ and $\sqrt{\hat B_{B_d}}F_{B_d}$, 
implying a smaller suppression of $\vtd$ than found in fig.~\ref{ckmplots}.
Similar comments can be made in connection with other quantities in 
table~\ref{mfv}. 

Clearly, low energy non-perturbative parameters as 
$\sqrt{\hat B_{B_d}}F_{B_d}$, $\hat B_K$ and $\xi$ do not depend on 
the compactification scale. On the other hand they are subject to 
uncertainties and
it is not surprising that the fit of the unitarity triangle within 
the ACD model prefers lower values of $\sqrt{\hat B_{B_d}}F_{B_d}$ and 
$\hat B_K$ than in the SM. The true effects of the KK modes can then 
only be clearly seen as in fig.~\ref{ckmplots} when the input parameters 
have no uncertainties. This exercise shows very clearly the importance 
of the reduction of the theoretical uncertainties in connection with 
the search for new physics.

\begin{table}[htb!]
\begin{center}
\begin{tabular}{|c|c|c|}
\hline
  Strategy       &         ACD ($1/R=200\gev$)   &            SM            \\
\hline  
$\bar {\eta}$  &         0.342 $\pm$ 0.027      &     0.357 $\pm$ 0.027   \\ 
                 &            $(0.288-0.398)$   &      $(0.305-0.411)$   \\
  $\bar {\rho}$  &         0.197 $\pm$ 0.047      &     0.173 $\pm$ 0.046    \\
                 &           $(0.102-0.296)$    &     $(0.076-0.260)$   \\
  $\sin 2\alpha$ &       $- 0.23 \pm 0.25$      &   $-0.09 \pm 0.25$       \\
                 &   $(-0.70-0.27)$                 & $(-0.54-0.40)$     \\
  $\gamma$       &         59.5 $\pm$ 7.0         &    63.5 $\pm$ 7.0        \\
  (degrees)      &   $(45.3-74.8)$                  & $(51.0-79.0)$       \\
  $\Delta M_s$   &         18.6$^{+1.9}_{-1.5}$   &   18.0$^{+1.7}_{-1.5}$ \\
  ($ps^{-1}$)    &   $(15.7-26.2)$                  & $(15.4-21.7)$       \\
$\vtd~(10^{-3})$ &          7.80 $\pm$ 0.42       &        8.15 $\pm$ 0.41   \\
                 &   $(6.96-8.69)$                &      $(7.34-8.97)$       \\
  \hline
\end{tabular}
\caption[]{ \small { Values and errors for different quantities in the ACD 
model with $1/R=200\gev$ \cite{PAST} and in the SM from \cite{BUPAST}. 
The 95$\%$  probability regions are  given in brackets.
}}
\label{mfv}
\end{center}
\end{table}

This discussion makes it clear that it is impossible to claim, as done 
in~\cite{CHHUKU}, that the reduction of the error on the parameter 
$\sqrt{\hat B_{B_d}}F_{B_d}$ by a factor of three will necessarily increase
the lowest allowed value of the compactification scale $1/R$. The lower 
bound on $1/R$ from $\Delta M_d$ depends necessarily on the actual value 
of $\sqrt{\hat B_{B_d}}F_{B_d}$. With decreasing $\sqrt{\hat B_{B_d}}F_{B_d}$
the lower bound on $1/R$ becomes weaker. One can easily check that 
decreasing the central value for $\sqrt{\hat B_{B_d}}F_{B_d}$ to $200\mev$, 
still within the present uncertainties, no significant lower bound on 
$1/R$ can be obtained even if the error on $\sqrt{\hat B_{B_d}}F_{B_d}$ is 
decreased by a factor of three.

\begin{figure}[hbt]
\centering
\includegraphics[scale=1]{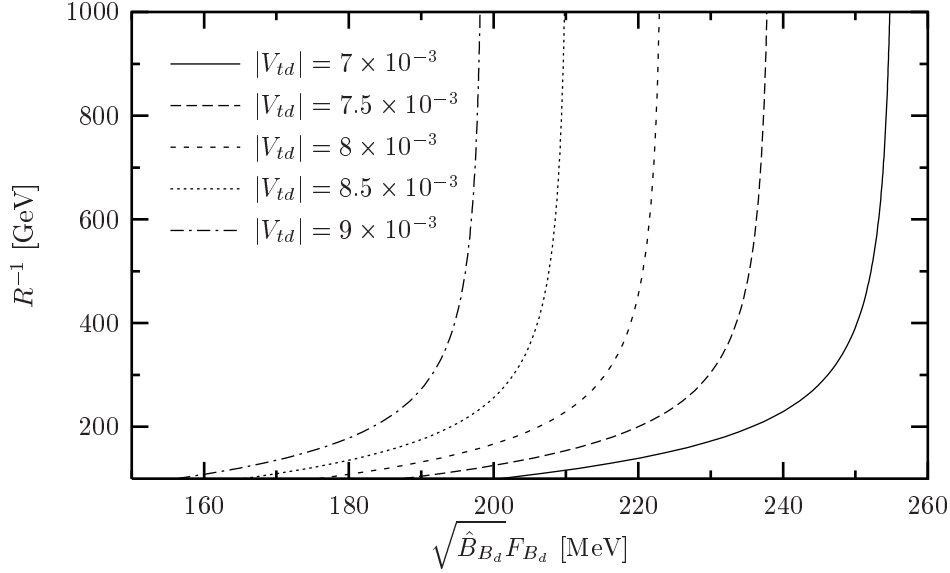} 
\caption[]{\small\label{fig:INVdeltaMd.eps} 
Correlation between $1/R$ and $\sqrt{\hat B_{B_d}}F_{B_d}$ for different values 
of $\vtd$.}
\end{figure}

In order to illustrate this point we show in fig.~\ref{fig:INVdeltaMd.eps} the correlation 
between $1/R$ and $\sqrt{\hat B_{B_d}}F_{B_d}$ for different values 
of $\vtd$ and $m_t=167\gev$. To this end we have used~(\ref{DMD}) with $\Delta M_d=0.503/ps$.
Clearly on the basis of $\Delta M_d$ alone it
is impossible to say anything about $1/R$. However, even if $\vtd$ is 
determined through $\Delta M_d/\Delta M_s$, the values of $1/R$ depend 
sensitively on $\sqrt{\hat B_{B_d}}F_{B_d}$ with the maximal and minimal 
values of $\sqrt{\hat B_{B_d}}F_{B_d}$ corresponding to maximal and minimal 
values of $1/R$, respectively. Thus a significantly improved lower bound on 
$\sqrt{\hat B_{B_d}}F_{B_d}$ could in principle provide an improved lower 
bound on $1/R$. However, if future lattice calculations will find 
values for $\sqrt{\hat B_{B_d}}F_{B_d}$ that are smaller than the present 
central value, $1/R$ could be forced to be low in order  for the ACD model
to fit the value of $\Delta M_d$.

\section{Rare K and B Decays}
\setcounter{equation}{0}

\subsection{Preliminaries}
We will now move to discuss
the semileptonic rare FCNC
transitions $\kpn$, $K_{\rm L}\to\pi^0\nu\bar\nu$, $K_L\to \mu\bar\mu$, 
$B\to X_{s,d}\nu\bar\nu$ and $B_{s,d}\to \mu\bar\mu$.
Within the SM and the ACD model these decays are loop-induced
semileptonic FCNC processes governed 
by $Z^0$-penguin and box diagrams and described by two functions 
$X(x_t,1/R)$ and $Y(x_t,1/R)$ for the decays with $\nu\bar\nu$ and 
$\mu\bar\mu$ in the final state, respectively.

A particular and very important virtue of these decays (with the 
exception of $K_L\to\mu\bar\mu$)
is their clean theoretical character \cite{Erice} that allows
to probe high energy scales of the theory and in particular
to measure $V_{td}$ and $\IM\lambda_t= \IM V^*_{ts} V_{td}$
from $K^+\to\pi^+\nu\bar\nu$ and $K_{\rm L}\to\pi^0\nu\bar\nu$
respectively.
Moreover, the combination of these two decays offers one of the
cleanest measurements of $\sin 2\beta$ \cite{BB4}, see \cite{BB4,GRNI} 
for more details.

\subsection{Effective Hamiltonians for \boldmath{$K\to\pi\nu\bar\nu$} and 
\boldmath{$B\to X_s\nu\bar\nu$}}
The effective Hamiltonian for $\kpn$  is given in the ACD model as
follows
\begin{equation}\label{hkpn} 
{\cal H}_{\rm eff}={G_{\rm F} \over{\sqrt 2}}{\alpha\over 2\pi 
\sin^2\theta_{ w}}
 \sum_{l=e,\mu,\tau}\left( V^{\ast}_{cs}V_{cd} X^l_{\rm NL}+
V^{\ast}_{ts}V_{td} \eta_X X(x_t,1/R)\right)
 (\bar sd)_{V-A}(\bar\nu_l\nu_l)_{V-A} \, .
\end{equation}
The index $l$=$e$, $\mu$, $\tau$ denotes the lepton flavour.
The dependence on the charged lepton mass resulting from the box diagram
is negligible for the top contribution. In the charm sector this is the
case only for the electron and the muon but not for the $\tau$-lepton.
In what follows we will set the QCD factor $\eta_X$ \cite{BB2,MU98,BB98}
to unity as for $\mt\equiv \overline m_t(\mt)$ it equals $0.994$.

The function $X(x_t,1/R)$ relevant for the top part is given by
\begin{equation}\label{xx9} 
X(x_t,1/R)=C(x_t,1/R)+B^{\nu\bar\nu}(x_t,1/R). 
\end{equation}
Here
\be\label{CACD}
C(x_t,1/R)=C_0(x_t)+\sum_{n=1}^\infty C_n(x_t,x_n)
\ee
results from $Z^0$-penguin diagrams with the SM contribution
$C_0(x_t)$ given by
\begin{equation}\label{C0xt}
C_0(x_t)={x_t\over 8}\left[{{x_t-6}\over{x_t-1}}+{{3x_t+2}
\over{(x_t-1)^2}}\;\ln x_t\right]~. 
\end{equation}
The sum in (\ref{CACD}) represents the KK contributions that are
calculated from the Feynman diagrams in fig. \ref{penguindiagrams} as discussed below.
Next
\be\label{BACD}
B^{\nu\bar\nu}(x_t,1/R)=-4 B_0(x_t)+
\sum_{n=1}^\infty B^{\nu\bar\nu}_n(x_t,x_n)
\ee
results from box diagrams with the SM contribution given by the first 
term and 
\begin{equation}\label{BF}
B_0(x_t)={1\over
4}\left[{x_t\over{1-x_t}}+{{x_t\ln x_t}\over{(x_t-1)^2}}\right].
\end{equation}
The sum in (\ref{BACD}) represents the KK contributions that are
calculated from the Feynman diagrams in fig. \ref{boxmumu} as discussed
below.

The expression corresponding to $X(x_t,1/R)$ in the charm sector is the 
function
$X^l_{\rm NL}$. It results from the NLO calculation \cite{BB3} and is given
explicitly in \cite{BB98} where further details can be found.
As in the case of the charm contributions in the $\Delta S=2$ Hamiltonian, 
 here the KK contributions are also negligible. 
The
numerical values for $X^l_{\rm NL}$ for $\mu = \mc$ and several values of
$\Lms^{(4)}$ and $\mc(\mc)$ can be found in \cite{BB98}. 
For our purposes we need only ($\lambda=0.221$)
\begin{equation}\label{p0k}
P_0(X)=\frac{1}{\lambda^4}\left[\frac{2}{3} X^e_{\rm NL}+\frac{1}{3}
 X^\tau_{\rm NL}\right]=0.41\pm0.06~,
\end{equation}
where the error results from the variation of $\Lms^{(4)}$ and $\mc(\mc)$.

In the case of $\klpn$ that is governed by CP-violating contributions 
only the top contribution in (\ref{hkpn}) matters. Similarly the 
effective Hamiltonians for $B\to X_{s,d}\nu\bar\nu$ are obtained from
(\ref{hkpn}) by neglecting the charm contribution and changing 
appropriately the CKM factor and quark flavours. 
In all these decays the KK modes 
contribute universally only through the function $X(x_t,1/R)$.

\subsection{Effective Hamiltonians for \boldmath{$B_{s,d}\to\mu\bar\mu$} and 
\boldmath{$K_L\to \mu\bar\mu$}}

The effective Hamiltonian for $B_s\to l^+l^-$ in the ACD model 
is given as follows:
\begin{equation}\label{hyll}
{\cal H}_{\rm eff} = -{G_{\rm F}\over \sqrt 2} {\alpha \over
2\pi \sin^2 \theta_{w}} V^\ast_{tb} V_{ts}
\eta_Y Y (x_t,1/R) (\bar bs)_{V-A} (\bar ll)_{V-A} + h.c.   \end{equation}
with $s$ replaced by $d$ in the
case of $B_d\to l^+l^-$.
The charm contributions are fully negligible
here. In what follows we will set the QCD factor $\eta_Y$ \cite{BB2,MU98,BB98}
to unity as for $\mt\equiv \overline m_t(\mt)$ it equals $1.012$.

The function $Y(x_t,1/R)$ is given in the ACD model by
\begin{equation}\label{yyx}
Y(x_t,1/R)  =C(x_t,1/R)+B^{\mu\bar\mu}(x_t,1/R),
\end{equation}
with $C(x_t,1/R)$ given in (\ref{CACD}) and
\be\label{BACDMU}
B^{\mu\bar\mu}(x_t,1/R)=-B_0(x_t)+
\sum_{n=1}^\infty B^{\mu\bar\mu}_n(x_t,x_n)
\ee
with the first term representing the SM contribution. The second term 
results from the box diagrams in fig. \ref{boxmumu}.

The effective Hamiltonian for the short distance contribution to 
$K_L\to\mu\bar\mu$ is given by (\ref{hyll}) with the appropriate change 
of the CKM factors and quark flavours and a small SM charm contribution 
$Y_{\rm NL}$ 
\cite{BB98} in analogy to 
$X^l_{\rm NL}$ in (\ref{hkpn}). The relevant branching ratio will be given 
below.

\subsection{\boldmath{$Z^0$}-Penguin Diagrams}
\begin{figure}[hbt]

  \centering
  \subfigure[]{
    \includegraphics[scale=0.9]{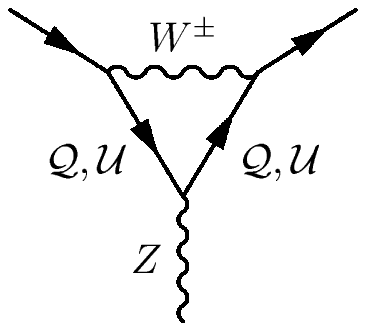}
    }
  \hspace{0.3cm}
  \subfigure[]{
    \includegraphics[scale=0.9]{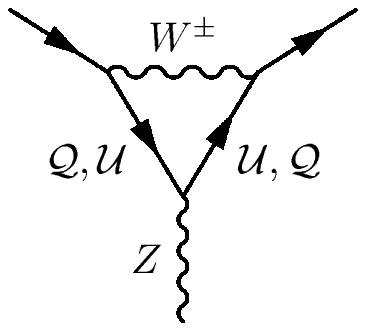}
    }
  \hspace{0.3cm}
  \subfigure[]{
    \includegraphics[scale=0.9]{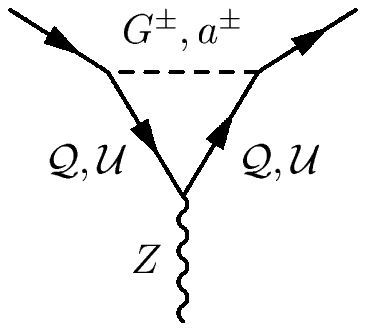}
    }
  \hspace{0.3cm}
  \subfigure[]{
    \includegraphics[scale=0.9]{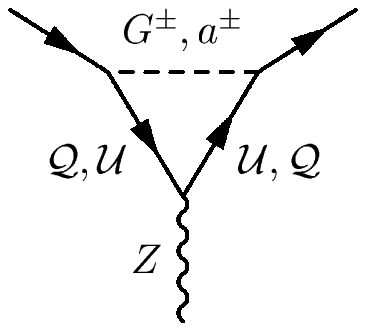}
    }
 \hspace{0.3cm}
  \subfigure[]{
    \includegraphics[scale=0.9]{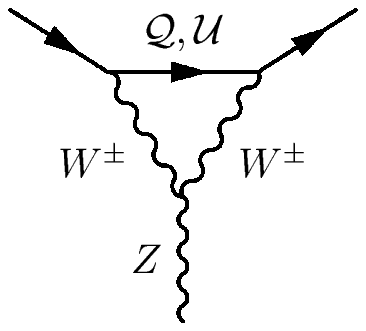}
    }
  \hspace{0.3cm}
  \subfigure[]{
    \includegraphics[scale=0.9]{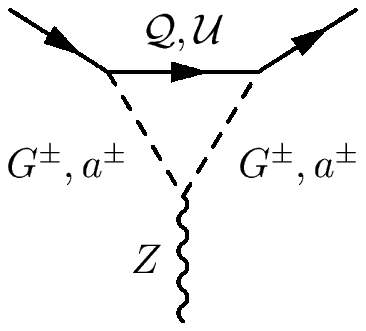}
    }
  \hspace{0.3cm}
  \subfigure[]{
    \includegraphics[scale=0.9]{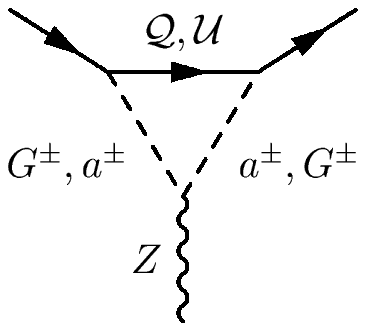}
    }
  \hspace{0.3cm}
  \subfigure[]{
    \includegraphics[scale=0.9]{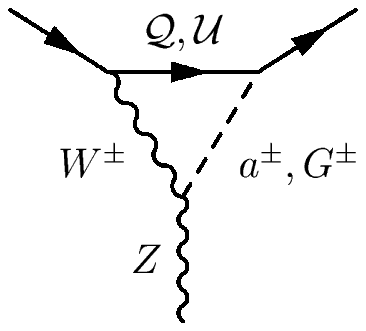}
    }
  \caption[]{\small\label{penguindiagrams} Penguin diagrams contributing to
  $C_n(x_t,x_n)$. The analytical expressions are listed 
in appendix \ref{apppenguin}. $(8)$ includes the additional diagram
with  $W^\pm$ and $(a^\pm,G^\pm)$ interchanged.}
\end{figure}

The function $C_n(x_t,x_n)$ can be found by calculating the vertex diagrams 
in fig.~\ref{penguindiagrams} and adding an electroweak counter term
as discussed in detail in \cite{BB1}. 
The latter is found by calculating the self-energy diagrams 
of fig.~\ref{selfenergydiagrams}  that describe flavour non-diagonal 
propagation of quark fields 
and subsequently rotating the quark fields appropriately so that this 
flavour non-diagonal
propagation  does not take place in the new fields.

In the 't Hooft-Feynman gauge for the $W_{(n)}^\pm$ and $G_{(n)}^\pm$  
propagators, the contribution of the diagrams in fig. \ref{penguindiagrams} to the 
flavour-changing vertex $\Gamma_Z^\mu$ including the electroweak counterterm 
is given by
\be\label{ZV}
\Delta \Gamma_Z^\mu=i \frac{g_2^3}{16\pi^2\cos\theta_w} V_{ts}^*V_{td}
                    C_n(x_t,x_n) \bar s \gamma_\mu (1-\gamma_5) d.
\ee
We neglect the external momenta in fig.~\ref{penguindiagrams} and 
the masses of 
external quarks.
The function $C_n(x_t,x_n)$ is defined through
\be\label{CN}
C_n(x_t,x_n)= F(x_{t(n)})-F(x_{u(n)})
\ee
with the functions
$F(x_{t(n)})$ and $F(x_{u(n)})$ representing the contributions of the 
$\mcQ\kkx tn$, $\mcU\kkx tn$ and $\mcQ\kkx un$, $\mcU\kkx un$ modes,
respectively,
\be
 F(x_{t(n)}) = \sum_{i=1}^8 F_i(x_{t(n)}) + \left(\frac12 - \frac13
s_w^2 \right) \sum_{i=1}^2 \Delta S_i ( x_{t(n)}).
\ee
Here $s_w\equiv\sin\theta_w$, $F_i$ stand for the contributions of 
diagrams in fig.~\ref{penguindiagrams} and $\Delta S_i$ denote the electroweak 
counter terms.

\begin{figure}[]
\renewcommand{\thesubfigure}{\space(\alph{subfigure})} 
\centering  
   \subfigure[]{\label{cncon.eps}
        \includegraphics[scale=.90]{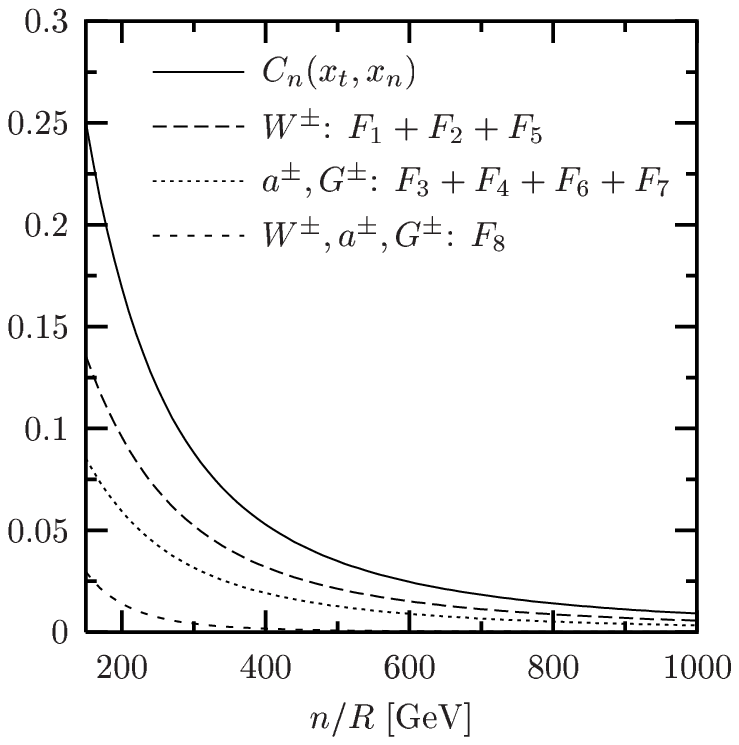} }\hspace{0.4cm}
   \subfigure[]{\label{csum.eps}
        \includegraphics[scale=.90]{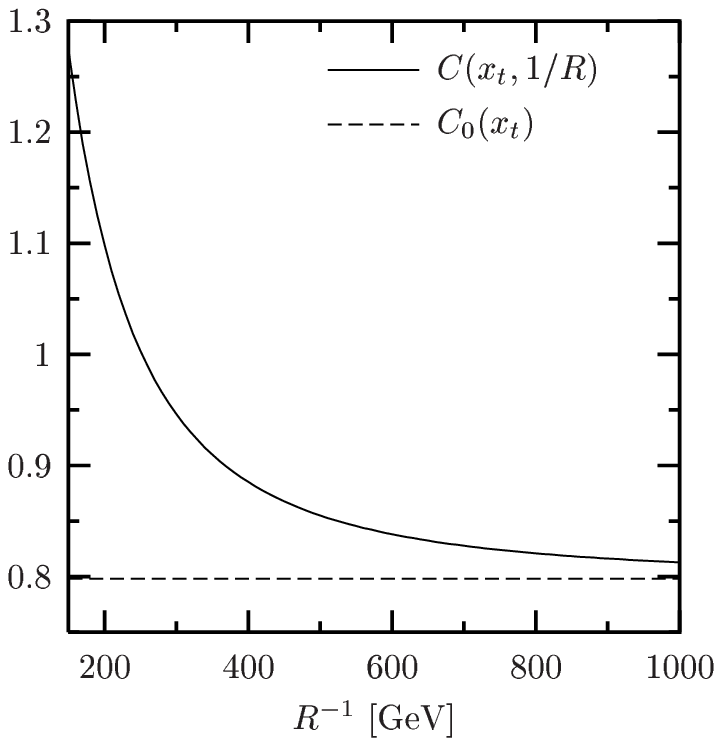} }
    \caption[]{\small\label{cnplot} (a) Contribution $C_n$ of the $n^{\rm th}$ KK mode to
  $C(x_t,1/R)$. The functions $F_i$ correspond to the diagrams given
  in fig.~\ref{penguindiagrams}.
  (b) The functions 
  $C(x_t,1/R)$ and $C_0(x_t)$. }
  \end{figure}

The explicit contributions of  
various sets of diagrams to these functions are given in appendix C. 
Adding up these contributions
we find
\be\label{CFIN}
C_n(x_t,x_n)=\frac{x_t}{8 (x_t-1)^2}
\left[x_t^2-8 x_t+7+(3 +3 x_t+7 x_n-x_t
x_n)\ln \frac{x_t+x_n}{1+x_n}\right]. 
\ee
In fig.~\ref{cncon.eps}, we show $C_n(x_t,x_n)$ as a function of $n/R$.
In this case the convergence of the sum of the KK modes is 
significantly improved by the GIM mechanism so that only a few first terms in
the sum in (\ref{CACD}) are relevant. We will return to this at 
the end of this section. 
We observe that in contrast to the function $S$ of section 3 the diagrams 
involving only $W^\pm_{(n)}$ play the dominant role among the KK contributions.

In fig.~\ref{csum.eps} we plot $C(x_t,1/R)$ versus 
$1/R$. 
For $1/R=200\gev$ we observe 
a $38\%$ enhancement of the function $C$ with respect to its SM value given 
by $C_0(x_t)$. 
For $1/R=250\gev$ this enhancement decreases to $26\%$ and it is $11\%$ for 
$1/R=400\gev$. The significant 
enhancement of $C$ is the origin of the enhancements of the branching ratios 
discussed below.

\subsection{\boldmath{$Z^0 b\bar b$} Vertex in the large \boldmath{$m_t$} limit}
The result in~(\ref{CFIN}) can also be used to find the dominant KK contribution to 
the $Z^0 b\bar b$ vertex in the large $m_t$ limit. As discussed already 
in~\cite{BB1}, in general the calculation of a low energy effective flavour 
violating $Z^0 d\bar s$ vertex cannot be directly compared with the full 
calculation of the $Z^0 b\bar b$ vertex. However, as pointed out there
 in the special limiting case
$m_t\gg M_W$ the two approaches, the direct evaluation of on-shell diagrams 
and the operator product expansion considered by us, are equivalent.
The reason for this is that in the limit $m_t\gg M_W$ with $1/R\ge m_t$ 
all the other mass scales as $m_b$ and external momenta of $b$-quarks 
become negligible compared with $m_t$ and $1/R$. Consequently with 
the definition
\be\label{Zbb}
\Gamma^\mu_{Zb\bar b}= i\frac{g_2}{\cos\theta_w} 
\bar b \gamma^\mu(g_L P_L+g_R P_R)b 
\ee
the one-loop contributions to the coupling $g_L$ can be found in the 
large $m_t$ limit by using the relation~\cite{BB1}
\be\label{bubu}
\delta g_L= \frac{G_F}{\sqrt{2}}\frac{M_W^2}{\pi^2} C(x_t,1/R).
\ee
In the case of the SM contributions this relation has been already 
analyzed in detail in~\cite{BB1} including  $\alpha_s$ to the one-loop 
$Z^0b \bar b$ vertex. For the KK contribution we find from~(\ref{CFIN})
\begin{eqnarray}\label{csummed}
\Delta C\!=\!\sum_{n=1}^{\infty} C_n ( x_t,x_n)
\!=\!\left( \frac{m_t^4 {\pi }^2}{96\,m_W^2} + \frac{m_t^2 {\pi }^2}{96} \right)  R^2 + 
  \left( \frac{m_t^4 {\pi }^4}{4320} - \frac{m_t^6 {\pi }^4}{2160\,m_W^2} - 
     \frac{m_t^2 m_W^2 {\pi }^4}{864} \right)  R^4 + \ldots\nnl
\end{eqnarray}
with the dominant contribution represented by the first term,  
 see appendix~\ref{CSUMAPP} for the derivation.
Retaining only this term and using~(\ref{bubu}) we find
\be\label{gL}
\delta g_L^{KK}=\frac{G_F}{\sqrt{2}} m_t^4\frac{R^2}{96} 
\ee
that agrees with a recent direct calculation of the KK contributiond to 
the $Zb\bar b$ vertex in~\cite{oliver:2002}, see their formula (3.6). Interestingly, 
while the corrections to the asymptotic result~(\ref{gL}) relevant 
for the $Zb\bar b$ vertex have been found in~\cite{oliver:2002} to be substantial, 
the result
\be
\Delta C=\frac{m_t^4\pi^2}{96 M_W^2} R^2
\ee
give a good approximation to the full KK contribution to the function 
$C(x_t,1/R)$ even for $m_t\approx 167\gev$.

\subsection{Box Diagram Contributions}
The functions $B^{\nu\bar\nu}_n(x_t,x_n)$ and $B^{\mu\bar\mu}_n(x_t,x_n)$
can be found by appropriate rescaling of the box diagrams contributing 
to $\Delta F=2$ transitions that we considered in section 3.  
It turns out that the box contributions 
of the KK modes are tiny. For instance for $1/R=200\gev$ and $\mt=167\gev$
we find $B^{\nu\bar\nu}_1(x_t,x_1)=0.0098$ and 
$B^{\mu\bar\mu}_1(x_t,x_1)=0.0049$ with even smaller values for $n>1$ and 
larger $1/R$. Consequently,
these contributions can be safely neglected in comparision with $C_n$. For
completeness we give in appendix D the analytic formulae for 
$B^{\nu\bar\nu}_n(x_t,x_n)$ and $B^{\mu\bar\mu}_n(x_t,x_n)$.
\begin{figure}[hbt]
  \begin{minipage}[b]{0.5\linewidth}
    \centering
    \includegraphics[scale=.9]{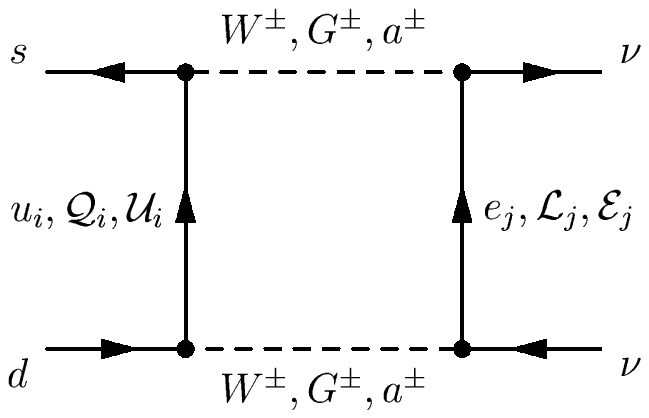}
  \end{minipage}
 \begin{minipage}[b]{0.5\linewidth}
    \centering
   \includegraphics[scale=.9]{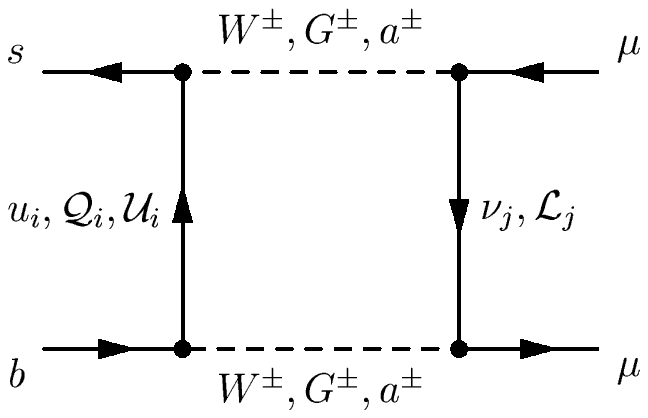}
 \end{minipage}
    \caption[]{\small\label{boxmumu} Box diagrams contributing to
      $B^{\nu\bar\nu}_n$ and to $B^{\mu\bar\mu}_n$.} 
  \end{figure}

\subsection{The Functions X and Y}
Neglecting the box contributions of the KK modes we find
\be\label{XACD}
X(x_t,1/R)=X_0(x_t)+\sum_{n=1}^\infty C_n(x_t,x_n)\equiv
X_0(x_t)+\Delta X~,
\ee

\be\label{YACD}
Y(x_t,1/R)=Y_0(x_t)+\sum_{n=1}^\infty C_n(x_t,x_n)\equiv
Y_0(x_t)+\Delta Y
\ee
with ($\mt=167~\gev$)
\begin{equation}\label{X0}
X_0(x_t)={{x_t}\over{8}}\;\left[{{x_t+2}\over{x_t-1}} 
+ {{3 x_t-6}\over{(x_t -1)^2}}\; \ln x_t\right]= 1.526 ~,
\end{equation}
\begin{equation}\label{Y0}
Y_0(x_t)={{x_t}\over8}\; \left[{{x_t -4}\over{x_t-1}} 
+ {{3 x_t}\over{(x_t -1)^2}} \ln x_t\right]= 0.980
\end{equation}
summarizing the SM contributions and $\Delta X=\Delta Y$ representing the
corrections due to KK modes. 
\begin{figure}[]
\renewcommand{\thesubfigure}{\space(\alph{subfigure})} 
\centering  
   \subfigure[]{\label{XandX0.eps}
        \includegraphics[scale=.90]{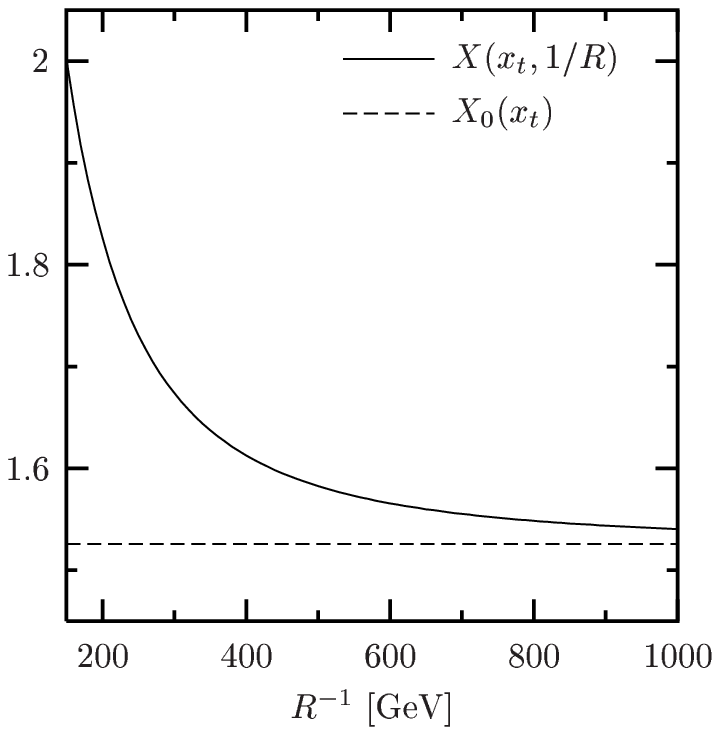} }\hspace{0.4cm}
   \subfigure[]{\label{YandY0.eps}
        \includegraphics[scale=.90]{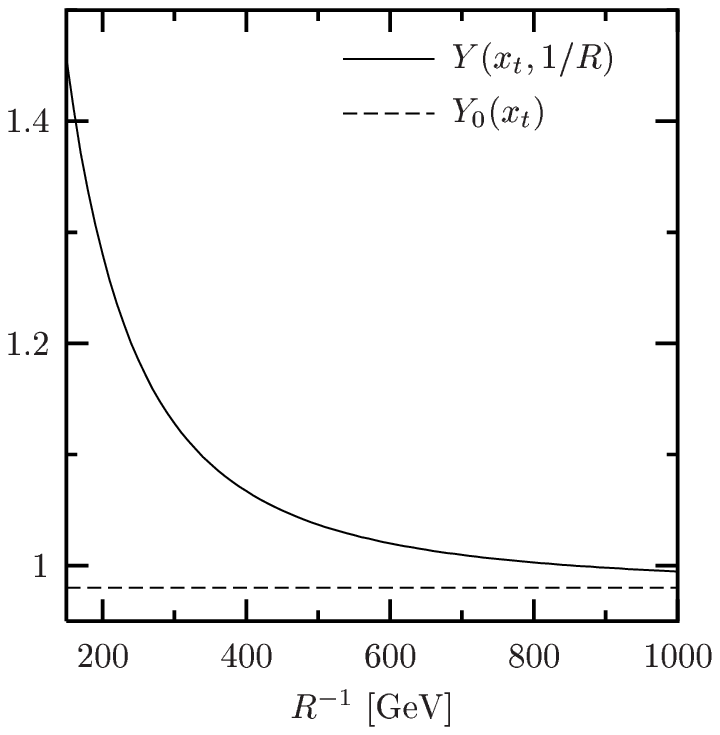} }
    \caption[]{\small\label{XandX0plot} The functions (a) $X(x_t,1/R)$
    and $X_0(x_t)$ and (b) $Y(x_t,1/R)$ and $Y_0(x_t)$.}
  \end{figure}

In fig.~\ref{XandX0plot} we plot $X(x_t,1/R)$ and $Y(x_t,1/R)$ versus $1/R$. 
 We observe that due to the inequality $X_0>Y_0$ the 
relative impact of the KK modes is larger in the function $Y$.
For $1/R=200\gev$ the functions $X$ and $Y$ are enhanced by $20\%$ and 
$31\%$, respectively. 
For $1/R=250\gev$ this enhancement decrease to $13\%(21\%)$ and are only 
 $6\%(9\%)$ for $1/R=400\gev$.

In table~\ref{SCXY} we give the values of the functions 
$S$, $C$, $X$ and $Y$ for different $1/R$ and $m_t=167\gev$.

\begin{table}[hbt]
\begin{center}
\begin{tabular}{|c||c|c|c|c|}\hline
 $1/R~[{\rm GeV}]$  & {$S$} & {$C$} 
& {$X$} & $Y$
 \\ \hline
$ 200$ & $ 2.813 $ & $ 1.099  $ & $1.826 $ &
 $1.281$ \\ \hline
$ 250$ & $ 2.664 $ & $ 1.003  $ & $1.731 $ &
 $1.185$ \\ \hline
$300$  & $  2.582 $ & $ 0.946  $ & $1.674  $ &
$1.128$ \\ \hline
$400$  & $ 2.500  $ & $ 0.885 $  & $1.613 $ &
$ 1.067 $\\ \hline
SM     & $2.398$ &    $ 0.798  $ & $ 1.526 $ &
$0.980$ \\ \hline
\end{tabular}
\end{center}
\caption[]{\small Values for the functions $S$, $C$, $X$ and $Y$.
\label{SCXY}}
\end{table}

\subsection{Branching Ratios for Rare Decays}
The branching ratios for the rare decays in question can be directly obtained 
from \cite{Erice} by simply replacing the SM functions $X_0$ and $Y_0$ by 
$X(x_t,1/R)$ and $Y(x_t,1/R)$, respectively. We have 
\begin{equation}\label{bkpn}
Br(\kpn)=\kappa_+\cdot\left[\left({\imlt\over\lambda^5}X(x_t,1/R)\right)^2+
\left({\relc\over\lambda}P_0(X)+{\relt\over\lambda^5}X(x_t,1/R)\right)^2
\right]~,
\end{equation}
\begin{equation}\label{kapp}
\kappa_+=r_{K^+}{3\alpha^2 Br(K^+\to\pi^0e^+\nu)\over 2\pi^2
\sin^4\theta_{w}}
 \lambda^8=4.31\cdot 10^{-11}\,,
\end{equation}
where we have used \cite{PDG}
\begin{equation}\label{alsinbr}
\alpha=\frac{1}{129},\qquad \sin^2\theta_{w}=0.23, \qquad
Br(K^+\to\pi^0e^+\nu)=4.87\cdot 10^{-2}\,.
\end{equation}
Here $\lambda_i=V^\ast_{is}V_{id}$ with $\lambda_c$ being
real to a very high accuracy. $r_{K^+}=0.901$ summarizes isospin
breaking corrections \cite{MP} in relating $\kpn$ to $K^+\to\pi^0e^+\nu$.
$P_0(X)$ is given in (\ref{p0k}).

Next,
\begin{equation}\label{bklpn}
Br(K_{\rm L}\to\pi^0\nu\bar\nu)=\kappa_{\rm L}\cdot
\left({\imlt\over\lambda^5}X(x_t,1/R)\right)^2~,
\end{equation}
\begin{equation}\label{kapl}
\kappa_{\rm L}=\frac{r_{K_{\rm L}}}{r_{K^+}}
 {\tau(K_{\rm L})\over\tau(K^+)}\kappa_+ =1.88\cdot 10^{-10}
\end{equation}
with $\kappa_+$ given in (\ref{kapp}) and
$r_{K_{\rm L}}=0.944$ summarizing isospin
breaking corrections in relating $\klpn$ to $K^+\to\pi^0e^+\nu$
\cite{MP}.

Next,
normalizing to $Br(B\to X_c e\bar\nu)$ and summing over three neutrino 
flavours we find 
\begin{equation}\label{bbxnn}
\frac{Br(B\to X_s\nu\bar\nu)}{Br(B\to X_c e\bar\nu)}=
\frac{3 \alpha^2}{4\pi^2\sin^4\theta_{w}}
\frac{|V_{ts}|^2}{|V_{cb}|^2}\frac{X^2(x_t,1/R)}{f(z)}
\frac{\kappa(0)}{\kappa(z)}\,.
\end{equation}
Here $f(z)$ is the phase-space factor for $B\to X_c
e\bar\nu$ with $z=\mc^2/\mb^2$  and $\kappa(z)=0.88$ 
\cite{CM78,KIMM} is the
corresponding QCD correction. The
factor $\kappa(0)=0.83$ represents the QCD correction to the matrix element
of the $b\to s\nu\bar\nu$ transition due to virtual and bremsstrahlung
contributions.
In the case of $B\to X_d\nu\bar\nu$ one has to replace $V_{ts}$ by
$V_{td}$ which results in a decrease of the branching ratio by
roughly an order of magnitude. In our numerical calculations we set 
$f(z)=0.54$ and $Br(B\to X_c e\bar\nu)=0.104$.

Next, the branching ratio for $B_s\to l^+l^-$ is given by 
\begin{equation}\label{bbll}
Br(B_s\to l^+l^-)=\tau(B_s)\frac{G^2_{\rm F}}{\pi}
\left(\frac{\alpha}{4\pi\sin^2\theta_{w}}\right)^2 F^2_{B_s}m^2_l m_{B_s}
\sqrt{1-4\frac{m^2_l}{m^2_{B_s}}} |V^\ast_{tb}V_{ts}|^2 Y^2(x_t,1/R)
\end{equation}
where $F_{B_s}$ is the $B_s$ meson
decay constant. The formula for $Br(B_d\to l^+l^-)$ is obtained by replacing
$s$ by $d$. The relevant input parameters are \cite{lellouch}
\be\label{input}
F_{B_d}=(203^{+27}_{-34})~{\rm MeV}, \qquad 
F_{B_s}=(238\pm 31)~{\rm MeV}~.
\ee
We set also $\tau(B_s)=1.46~{\rm ps}$ and $\tau(B_d)=1.54~{\rm ps}$
 \cite{PDG}. 
The short distance contribution to the dispersive part of $\klm$ is given 
by \cite{BB2,BB3} 
\begin{equation}\label{bklm}
Br(\klm)_{\rm SD}=\kappa_\mu\left[\frac{\relc}{\lambda}P_0(Y)+
\frac{\relt}{\lambda^5} Y(x_t,1/R)\right]^2,
\end{equation}
\begin{equation}\label{kapm}
\kappa_\mu=\frac{\alpha^2 Br(K^+\to\mu^+\nu)}{\pi^2\sin^4\theta_{w}}
\frac{\tau(K_{\rm L})}{\tau(K^+)}\lambda^8=1.733\cdot 10^{-9}\,,
\end{equation}
where we have used $Br(K^+\to\mu^+\nu)=0.634$.
The charm contribution including NLO corrections is given by \cite{BB98}  
\begin{equation}\label{p0kldef}
P_0(Y)=\frac{Y_{\rm NL}}{\lambda^4} = 0.128\pm 0.013.
\end{equation}
Unfortunately due to long distance contributions to the dispersive part of 
$K_L\to \mu^+\mu^-$, the extraction of $Br(\klm)_{\rm SD}$ from the data is 
subject to considerable uncertainties. The present best estimate reads 
\cite{ISIDORI}
 \begin{equation}\label{SD}
Br(\klm)_{\rm SD}\le 2.5\cdot 10^{-9}~.
\end{equation}

\subsection{GIM Mechanism and Convergence of the KK Sum}
\label{GIMandConvergence}

The tree-level masses of the KK modes of all particles approach the value
$n/R$ for large KK mode numbers $n$, see
eq.~(\ref{fermionmassformula}) and~(\ref{GBmasses}). Due to the unitary 
CKM matrix the
GIM mechanism~\cite{GIM} comes into play. It suppresses partly the
higher KK mode contributions to the sums in
eq.~(\ref{SACD}),~(\ref{xx9}),~(\ref{BACD}) and~(\ref{BACDMU}) and is
essential for the determination of the dominant contributions. For
the $S_n$ function the GIM suppression amounts to an additional factor of
$1/n^4$ for the contributions ${WW}$, ${GG}$ and ${WG}$. The
contributions from diagrams with $a^\pm$ are not suppressed due to the
second term in the coupling $m_4^{(t)}$, see (\ref{mparameters}). 
This results in a hierarchy
of the various contributions to $S_n$ with ${WW}$, ${GG}$ and ${WG}$
proportional to $1/n^6$, ${Wa}$ and ${Ga}$ proportional to $1/n^4$
and the dominant contribution ${aa}$ proportional to $1/n^2$ for
large values of $n$.

For the penguin diagrams the effect of the GIM mechanism on the convergence
of the particular contributions is partly hidden by the subtle
cancellation of divergencies among the different contributions and the
two self-energy diagrams. 
For the combinations plotted in fig.~\ref{cnplot} we observe that the
term corresponding to $W^\pm: F_1+F_2+F_5$ is logarithmically
divergent without GIM and shows a $1/n^2$ behaviour after GIM mechanism 
has been taken into account.
The $a^\pm,G^\pm: F_3+F_4+F_6+F_7$ term shows no GIM
suppression and is proportional to $1/n^2$ for large values of
$n$. The mixed term $W^\pm,a^\pm,G^\pm: F_8$ is constant for large $n$ before
the inclusion of GIM mechanism, but is
GIM suppressed by a factor of $1/n^2$.

The contributions to
$B^{\nu\bar\nu}_n$ and $B^{\mu\bar\mu}_n$ show a $1/n^4$ behaviour
with and without GIM, except the $G_{WW}$ and $H_{WW}$
terms. They are proportional to $1/n^2$ without GIM and behave like
$1/n^4$ with GIM suppression at work. This different asymptotic
behaviour with respect to the $S_n$
function is due to the appearance of
leptons with negligible zero-mode masses instead of quarks in 
the box diagrams contribution to the functions 
$G$ and $H$, see eq.~(\ref{mparameters}).

\subsection{Numerical Analysis}
As discussed in section 3, $\vtd$, $\bar\varrho$ and $\bar\eta$  in the SM and
in the ACD model differ from each other. In our numerical analysis we will 
take this difference into account. Not taking this difference into account 
would misrepresent significantly the patterns of the enhancements of 
branching ratios in question.

\begin{figure}[hbt]
  \begin{minipage}[b]{0.5\linewidth}
    \centering
    \includegraphics[scale=.9]{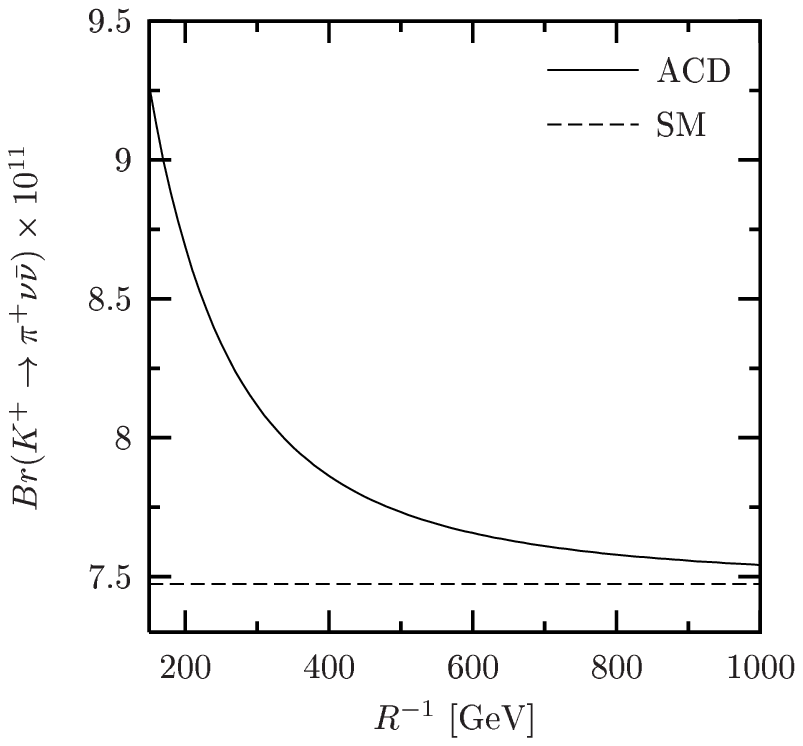}
  \end{minipage}
 \begin{minipage}[b]{0.5\linewidth}
    \centering
   \includegraphics[scale=.9]{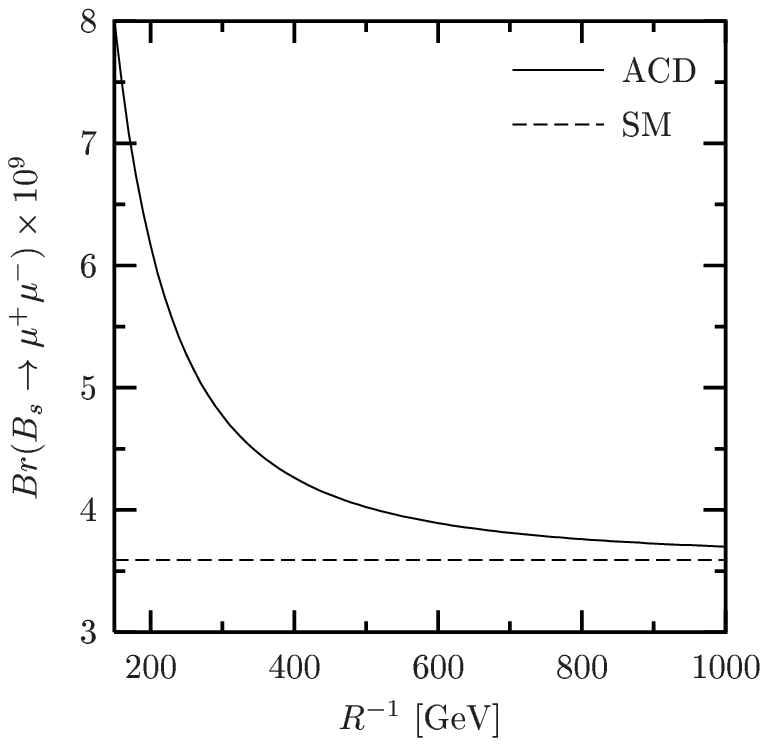}
\end{minipage}
  \caption[]{\small Branching ratio for the decays $K^+\rightarrow \pi^+ \nu
  \overline{\nu}$ and  $B_s\to\mu^+ \mu^-$ as predicted by the ACD model and 
  the SM as 
  functions of the inverse radius $R$ of the extra dimension.\label{BR} }
  \end{figure}

In fig.~\ref{BR} we show the branching ratios $Br(\kpnn)$ and 
$Br(B_s\to \mu^+\mu^-)$ as functions of the compactification scale $1/R$. 
As $\vts_{ACD}=\vts_{SM}$, the enhancement of $Br(B_s\to \mu^+\mu^-)$ is 
entirely governed by the ratio $(Y/Y_0)^2$. On 
the other hand the dependence of $Br(\kpnn)$ on $1/R$ differs from the one of
$(X/X_0)^2$ because of the additional charm contribution 
that is independent of $1/R$ and the fact that $V_{td}$ in the ACD model 
differs from its SM value. The remaining branching ratios are shown 
in table~\ref{brtable}. The $1/R$ dependence of $Br(B\to X_s\nu\bar\nu)$ 
is governed by 
the function $X^2(x_t/1/R)$, while the corresponding dependences of 
$Br(B\to X_d\nu\bar\nu)$, $Br(\klpn)$, $Br(B_d\to\mu^+\mu^-)$ and 
$Br(\kmm)_{\rm SD}$ include also the $1/R$ dependence of $V_{td}$ shown in 
fig.~\ref{Vtdplot}. As expected, all the branching ratios are significantly 
enhanced for $1/R\le 300\gev$. For $1/R\ge 400\gev$, except for
$Br(B_s\to \mu^+\mu^-)$,  the distinction between the predictions of the ACD 
model and the SM will be very difficult.

\begin{table}[hbt]

\vspace{0.4cm}
\begin{center}
\begin{tabular}{|c||c|c|c|c|c|}\hline
{$1/R$ } & {$200\gev$} & {$250\gev$}&  {$300\gev$} 
& {$400\gev$} & SM
 \\ \hline
$Br(\kpnn)\times 10^{11}$ & $ 8.70 $ & $8.36$ & $ 8.13$ & $ 7.88$ &  $7.49$ 
\\ \hline
$Br(\klpn)\times 10^{11}$ & $ 3.26 $ & $3.17$ & $ 3.09 $ & $ 2.98$ &  $2.80$ 
\\ \hline
$Br(\klm)_{\rm SD}\times 10^{9} $ & $ 1.10 $ & $1.00$ &  $ 0.95$ & $ 0.88$ &
$0.79$\\ \hline
$Br(B\to X_s\nu\bar\nu)\times 10^{5}$ & $5.09 $ & $4.56$ & $ 4.26 $ & $ 3.95$ &
$3.53$ \\ \hline
$Br(B\to X_d\nu\bar\nu)\times 10^{6}$ & $1.80 $ & $1.70$ & $ 1.64$ & $ 1.58$ &
$1.47$ \\ \hline
$Br(B_s\to \mu^+\mu^-)\times 10^{9}$ & $6.18 $ & $5.28$ & $ 4.78$ & $4.27$ &
$3.59$ \\ \hline
$Br(B_d\to \mu^+\mu^-)\times 10^{10}$ & $1.56$ & $1.41$ & $ 1.32 $ & $ 1.22$ &
$1.07$ \\ \hline
\end{tabular}
\caption[]{\small Branching ratios for rare decays in the ACD model and the 
SM as discussed in the text.
\label{brtable}}
\end{center}
\end{table}

In this numerical analysis we have used the results for the CKM parameters
of fig.~\ref{ckmplots} and the central values of all the remaining 
input parameters as given above. 
The uncertainties in these parameters  partly cover the 
differences between the ACD model and the SM model and it is essential 
to reduce these uncertainties considerably if one wants to see the effects 
of the KK modes in the branching ratios in question.
Therefore a detailed analysis that includes all uncertainties would be in our
opinion premature at present.

\subsection{An Upper Bound on \boldmath{$Br(\kpnn)$} in the ACD Model}
The enhancement of $Br(\kpnn)$ in the ACD model is interesting in view
of the results  from the AGS E787
collaboration at Brookhaven \cite{Adler01} that read 
\be\label{kp01}
Br(K^+ \rightarrow \pi^+ \nu \bar{\nu})=
(15.7^{+17.5}_{-8.2})\cdot 10^{-11}
\end{equation}
with the central value  by a factor of 2 above the SM expectation. Even if 
the errors are substantial and this result is compatible with the SM, 
the ACD model with a low compactification scale is  closer to the 
data. As emphasized in \cite{Adler01,AI01} the central value in (\ref{kp01})
implies within the SM a value for $\vtd$ that is substantially higher 
than the value obtained from the standard analysis of the UT of section 3.
Here we would like to emphasize that within the ACD model 
$Br(K^+ \rightarrow \pi^+ \nu \bar{\nu})$ is closer to the data in spite of 
the fact that $\vtd_{\rm ACD}< \vtd_{\rm SM}$.
The enhanced $Z^0$--vertex represented by the function $C$ is responsible for
this behaviour.

In \cite{BB98} an upper bound on $Br(K^+ \rightarrow \pi^+ \nu \bar{\nu})$
has been derived within the SM. This bound depends only on $\vcb$, $X_0$, 
$\xi$ and $\Delta M_d/\Delta M_s$. With the precise value for the angle 
$\beta$ now available this bound can be turned into a useful formula for 
$Br(K^+ \rightarrow \pi^+ \nu \bar{\nu})$ \cite{AI01} that expresses
this branching ratio in terms of theoretically clean observables. 
In the ACD model this formula reads:
\be\label{AIACD}
Br(K^+ \rightarrow \pi^+ \nu \bar{\nu})=
\bar\kappa_+\vcb^4 X^2(x_t,1/R)\left[ \sigma R^2_t\sin^2\beta+
\frac{1}{\sigma}\left(R_t\cos\beta+
\frac{\lambda^4P_0(X)}{\vcb^2X(x_t,1/R)}\right)^2\right],
\ee
where $\sigma=1/(1-\lambda^2/2)^2$, $\bar\kappa_+=\kappa_+/\lambda^8$ and 
$R_t$ is given in (\ref{Rt}).
This formula is theoretically very clean and does not involve 
hadronic uncertainties except for $\xi$ in (\ref{Rt}) and to a lesser 
extent in $\vcb$. 

In order to find the upper bound on $Br(\kpnn)$ in the ACD model we use
\be
\vcb\le 0.0422~,~\quad P_0(X)<0.47~,~\quad \sin\beta=0.40~, ~\quad 
\mt<172~\gev,
\ee
where we have set $\sin\beta$ to its central value (see (\ref{ga})) as 
$Br(\kpnn)$ depends very weakly on it. The bound on $\vcb$ results from 
$\vcb=0.0406\pm 0.0008$ \cite{BUPAST}. We used here two standard deviations 
as the determination of $\vcb$ involves some hadronic uncertainties.
The result of this exercise is shown in table~\ref{Bound}. We give 
there $Br(\kpn)_{\rm max}$  
as a function of $\xi$ and $1/R$ for two different values of $\Delta M_s$.
The range for $\xi$ chosen by us is in accordance with (\ref{lat}) and 
the recent new analysis in \cite{BEFAPRZU} that gives $\xi=1.22\pm 0.07$.
The upper bound in the SM given in the last column is lower than the values 
for $1/R=400~\gev$ by roughly $10\%$.
We observe that for $1/R=200~\gev$ and $\xi=1.30$ the maximal value
for $Br(\kpnn)$ in the ACD model is rather close to the central value in
(\ref{kp01}). 

At first sight the $30\%$ enhancement of $Br(\kpnn)$ for $1/R=200\gev$ 
with respect to the 
SM values seems to contradict the results in table~\ref{brtable}, where 
a more modest enhancement of $Br(\kpnn)$ is seen. However, one should 
realize that now $\Delta M_d/\Delta M_s$ and not $\Delta M_d$ alone enters 
the analysis and the enhancement of the function $X$ is not accompanied by 
a suppression of $\vtd$, that with $R_t$ given by (\ref{Rt}), equals the one 
in the SM. The consistency with the experimental value of $\Delta M_d$ 
requires then a sufficiently small $\sqrt{\hat B_{B_d}} F_{B_d}$. 
In table~\ref{Bound} we indicate by a star the cases that require  
$\sqrt{\hat B_{B_d}} F_{B_d}\le 190\mev$. Such low values are rather 
improbable 
in view of (\ref{lat}) and consequently values of $Br(\kpnn)$ larger than 
$12\cdot 10^{-11}$ are rather unlikely even in the ACD model.

\begin{table}[hbt]
\vspace{0.4cm}
\begin{center}
\begin{tabular}{|c||c|c|c|c|c|}\hline
{$\xi$ } & {$1/R=200~\gev$} & {$1/R=250~\gev$}& {$1/R=300~\gev$} 
& {$1/R=400~\gev$} & SM
 \\ \hline
$1.30$ & $ 13.8^*~(12.3^*) $ & $ 12.7^*~(11.3^*)$ & $ 12.0^*~(10.7)$ & 
$ 11.3^*~(10.1)$ &  $10.8~(9.3)$ \\ \hline
$1.25$ & $ 13.0^*~(11.6) $ & $ 12.0~(10.7) $ & $ 11.4~(10.2) $ & 
$ 10.7~(9.6)$ & $10.3~(8.8)$ \\ \hline
$1.20$ & $ 12.2^*~(10.9) $ & $ 11.3~(10.1) $ & $ 10.7~(9.6) $ & 
$ 10.1~(9.1)$ & $9.7~(8.4)$\\ \hline
$1.15$ & $11.5~(10.3) $ & $10.6~(9.5) $ & $ 10.1~(9.0) $ & $ 9.5~(8.5)$ &
$9.1~(7.9)$ \\ \hline
\end{tabular}
\caption[]{\small Upper bound on $Br(\kpnn)$ in units of $10^{-11}$ for 
different 
values of $\xi$ and $1/R$ and $\Delta M_s=18/{\rm ps}~(21/{\rm ps})$. The 
stars indicate the results corresponding to 
$\sqrt{\hat B_{B_d}} F_{B_d}\le 190\mev$.
\label{Bound}}
\end{center}
\end{table}

\section{Summary and Outlook}
In this paper we have calculated for the first time the contributions of 
the Kaluza-Klein (KK) modes to $\Delta M_K$, $\varepsilon_K$, $\Delta M_{d,s}$
 and rare decays 
$\kpn$, $\klpn$, $\klm$, $B\to X_{s,d}\nu\bar\nu$ and $B_{s,d}\to\mu\bar\mu$ 
in the Appelquist, Cheng and Dobrescu (ACD) model with one universal extra 
dimension. As a byproduct we have given a list of the required Feynman 
rules that have not been presented in the literature so far.

 The nice property of this extension of the SM is the presence of
only a single new parameter, $1/R$.
This economy in new parameters should be 
contrasted with supersymmetric theories and models with an extended Higgs 
sector.
Taking  $1/R=200~\gev$ our findings are as follows:
\begin{itemize}
\item
The short distance one-loop function $S(x_t,1/R)$ relevant for $\Delta F=2$ 
transitions is larger than the corresponding SM function $S_0(x_t)$ by 
roughly $17\%$. This implies on the basis of $\varepsilon_K$ and $\Delta M_d$
a $8\%$ suppression of $\vtd_{\rm ACD}$ with respect to $\vtd_{\rm SM}$
and a decrease of the angle $\gamma_{\rm ACD}$ by $10^\circ$ with respect to  
$\gamma_{\rm SM}$. On the other hand, 
$(\Delta M_s)_{\rm ACD}$ is larger than $(\Delta M_s)_{\rm SM}$ by  
$17\%$, see section 3 for details. $\Delta M_K$ is essentially 
uneffected.
\item
In order to see whether the modifications of the SM expectations are required
by the data, the comparision of $\vtd$ and $\gamma$ extracted from 
$\varepsilon_K$ and $\Delta M_d$ in the SM and in the ACD model with the 
universal unitarity triangle constructed by means of $\vub$, 
$\Delta M_d/\Delta M_s$ and $a(\psi K_S)$ will be important. To this end,
the data on these quantities have to be improved  and the uncertainties in 
the relevant non-perturbative parameters reduced.
\item
The short distance one-loop function $X(x_t,1/R)$ relevant for the decays 
$\kpn$, $\klpn$ and $B\to X_{s,d}\nu\bar\nu$ 
is larger than the corresponding SM function $X_0(x_t)$ by 
roughly $20\%$ due to KK contributions to the $Z^0$-penguins. 
In the case of $\kpn$, $\klpn$ and $B\to X_{d}\nu\bar\nu$ this enhancement is 
partially compensated by the fact that 
$\vtd_{\rm ACD}<\vtd_{\rm SM}$ and $\bar\eta_{\rm ACD}<\bar\eta_{\rm SM}$. 
We then find the enhancements of $Br(\kpn)$, $Br(\klpn)$ and 
$Br(B\to X_{d}\nu\bar\nu)$ over the 
SM expectations by $16\%$, $17\%$ and $22\%$, respectively. 
As $B\to X_s \nu\bar\nu$ 
is governed by the CKM element $\vts$ that is common to the SM and the ACD 
model, the enhancement of $Br(B\to X_s \nu\bar\nu)$ amounts to $44\%$.
\item
The short distance one-loop function $Y(x_t,1/R)$ relevant for the decays 
$B_{d,s}\to\mu\bar\mu$ and $K_L\to \mu\bar\mu$,  
is larger than the corresponding SM function $Y_0(x_t)$ by 
roughly $30\%$. As $B_{s}\to\mu\bar\mu$ is governed by the CKM element
$\vts$ with $\vts_{\rm ACD} = \vts_{\rm SM}$
this implies a $72\%$ enhancement of $Br(B_{s}\to\mu\bar\mu)$ relative to the 
SM expectation. In the case of $B_d\to \mu\bar\mu$ and $K_L\to\mu\bar\mu$, 
due to $\vtd_{\rm ACD}<\vtd_{\rm SM}$,
the corresponding enhancements amount to  $38\%$ and $46\%$.  
\item
As the ACD model belongs to the class of MFV models, general properties 
of these models identified in \cite{UUT,REL} are automatically valid here.
\item
For $1/R= 250~(300)~\gev$ all these
effects are decreased roughly by a  factor of $1.5~(2.0)$.
See fig.~\ref{BR} and table~\ref{brtable}.
\end{itemize}

In short, the signatures of the deviations from the SM expectations are:
\begin{itemize}
\item
The decrease of $\vtd$, $\gamma$, $\bar\eta$ and $\bar\varrho$.
\item
The increase of $\Delta M_s$.
\item
The increase of all branching ratios considered in this paper with 
a hierarchical structure of maximal enhancements: 
$\kpn~(16\%)$, $\klpn~(17\%)$, 
$B\to X_{d}\nu\bar\nu~(22\%)$, $K_L\to\mu\bar\mu~(38\%)$, 
 $B\to X_{s}\nu\bar\nu~(44\%)$, 
 $B_{d}\to\mu\bar\mu~(46\%)$ and 
$B_{s}\to\mu\bar\mu~(72\%)$ for $1/R=200\gev$. For $1/R= 250~(300)\gev$ 
 these enhancements are decreased roughly by a  factor of $1.5~(2.0)$.
\end{itemize}

In the coming years of particular interest will be the improved 
measurements of 
$Br(K^+\to\pi^+\nu\bar\nu)$ and $\Delta M_s$. Indeed, in the MFV models 
$Br(K^+\to\pi^+\nu\bar\nu)$ can be predicted as a function of $X$ once 
the angle $\beta$ and $\Delta M_d/\Delta M_s$ are known. 
The relevant formula is given in (\ref{AIACD}).
For a given value 
of $X$, 
the branching ratio $Br(K^+\to\pi^+\nu\bar\nu)$ decreases with increasing 
$\Delta M_s$. If the present central experimental values for 
$Br(K^+\to\pi^+\nu\bar\nu)$ will remain while the experimental error 
will decrease significantly, the SM expectations for 
$Br(K^+\to\pi^+\nu\bar\nu)$ will be significantly below the experimental 
data while the
corresponding estimates within the ACD model may agree with them due to
$X(x_t,1/R)>X_0(x_t)$, see table~\ref{Bound}.

Another definite prediction of the ACD model is an 
{\it increase} of $\Delta M_s$ over the SM value. This should be contrasted 
with the prediction of the MSSM at large $\tan\beta$ where a {\it suppression} 
of $\Delta M_s$ with respect to the SM value is predicted \cite{BUCHROSL}. 
Thus, when the 
ratio $(\Delta M_s)_{\rm exp}/(\Delta M_s)_{\rm SM}$ will be known with 
a sufficient accuracy, we will know whether the ACD model with a low $1/R$ or 
the MSSM with a 
large $\tan\beta$ is ruled out by the data.

A distinction between the ACD model and the MSSM at low $\tan\beta$
will also be possible. In the latter case the supersymmetric effects in 
$\Delta F=2$ transitions considered in section 3 are generally larger than
in $\Delta F=1$ transions so that the branching ratios for   
$\kpn$, $\klpn$, $\klm$, $B\to X_{d}\nu\bar\nu$ and $B_{d}\to\mu\bar\mu$
are generally suppressed with respect to the SM while they can be enhanced 
for special ranges of supersymmetric parameters in
the case of  $B\to X_{s}\nu\bar\nu$ and $B_{s}\to\mu\bar\mu$ 
\cite{BRMSSM}.

However, the main message from our analysis is the following one. 
Even for the lowest compactification scale, 
$1/R=200~\gev$, considered by us,
the ACD model is consistent with all the available data on FCNC processes 
analyzed here . No fine tunning of the 
parameters characteristic in the flavour sector for general 
supersymmetric models 
is necessary. This is first of all connected with the
GIM mechanism that assures the convergence of the sum over 
the KK modes in the case of $Z^0$ penguin diagrams, 
removing the sensitivity of the calculated
branching ratios to the scale $M_s\gg 1/R$ at which the higher dimensional 
theory becomes non-perturbative and at which the towers of the KK particles 
must be cut off in an appropriate way.

With the much improved data on the processes calculated in this 
paper and the theoretical uncertainties reduced, it should be possible 
in the future to 
distinguish the predictions of the ACD model from the SM ones, provided 
$1/R< 400\gev$. For higher compactification scales this distinction 
will be very difficult with a possible exception of $B_s\to \mu^+\mu^-$. 
Whether these findings apply also to
$B\to X_s\gamma$, $B\to X_s l^+l^-$ and $K_L\to \pi^0 e^+e^-$ is an 
interesting question. As the $Z^0$-penguins are enhanced in the ACD model, 
the corresponding enhancement of the branching ratios for
$B\to X_s l^+l^-$ and $K_L\to \pi^0 e^+e^-$ can easily be calculated by 
means of the known formulae \cite{Erice,BS98,BHI}. However, such an analysis 
would clearly be unsatisfactory as these decays receive also contributions 
from the $\gamma$-penguins and magnetic penguins. The KK contributions to
$\gamma$-penguins are unknown, while the corresponding contributions to 
the magnetic penguins have been calculated in the context of an analysis of 
the decay 
$B\to X_s\gamma$ under the assumption of the dominance of the scalars 
$a^\pm_{(n)}$ \cite{AGDEWU}. We have seen that this assumption would be 
correct in the case
of $S(x_t,1/R)$ for $1/R\ge 300\gev$, but certainly would misrepresent the 
KK contributions to $C(x_t,1/R)$ in view of strong cancellations between 
different contributions as discussed in section 4. Consequently a satisfactory
analysis of $B\to X_s\gamma$, $B\to X_s l^+l^-$ and $K_L\to \pi^0 e^+e^-$ 
requires a priori the inclusion of all KK contributions.
We  will address all these issues in the forthcoming paper \cite{BPSW02}. 

\section*{Acknowledgements}
We thank  Fabrizio Parodi and  Achille Stocchi for providing the numbers 
in the second 
 column of table~\ref{mfv} and Geraldine Servant for interesting comments.
This
research was partially supported by the German `Bundesministerium f\"ur 
Bildung und Forschung' under contract 05HT1WOA3 and by the 
`Deutsche Forschungsgemeinschaft' (DFG) under contract Bu.706/1-1.

\appendix
\pagebreak

\newsection{Feynman Rules in the ACD Model}\label{Feynmanrules}
\allowdisplaybreaks[1]

In this section, we list all propagators and vertex rules needed for
the calculation of the box and penguin diagrams considered in this paper.
The Feynman rules are derived in the 5d $R_\xi$-gauge described in
Section~\ref{gaugefixing}.

\label{couplingconversion}
It is convenient to define a 4 dimensional counterpart to every
parameter in the 5 dimensional Lagrangian, marked there with a
caret. The conversion factors are chosen in order to eliminate all
explicit appearances of factors of $\sqrt{2\pi R}$ in the Feynman
rules: 
\be
  \begin{alignedat}{3}
    \mu &= \hat\mu, &\hspace{14ex}
    v &= \sqrt{2 \pi R}~\hat v, &\hspace{14ex}
    g_2 &= \frac{1}{\sqrt{2 \pi R}}~\hat g_2,\\
    g' &= \frac{1}{\sqrt{2\pi R}}~\hat g', &
    \lambda_{\mcU / \mcD} &= \frac{1}{\sqrt{2\pi R}}~\hat\lambda_{\mcU
      / \mcD},
  \end{alignedat}
\ee
where $\hat\mu$ is the Higgs mass parameter.

In order to simplify the notation, we omit the KK indices of the fields and
of the mass parameters defined in (\ref{mparameters}) and
(\ref{Mparameters}). There is no 
ambiguity because in one-loop calculations at least one field is
always a zero-mode. In our case, this is the $Z$ boson in the $Z$
vertices, and the down-type quark or the neutrino in all other 
vertices. Due to KK parity conservation, the other two
fields have equal KK mode number, i.e. either zero or $n \geq 1$.

Fermion zero-modes have substantially different Feynman rules than
their KK excitations.
The fields $\mcQ_u$, $\mcQ_d$, $\mcU$, $\mcD$, $\mcL_\nu$, $\mcL_e$
and $\mcE$ are always supposed to be $(n\geq 1)$-modes, while the
zero-modes are labeled $u$, $d$, $\nu$ and $e$.
The generation indices are $i=u,c,t$ and $j=d,s,b$ for the quarks and
$i,j=e,\mu,\tau$ for the leptons.

The masses of all bosonic particles can
be expressed in terms of four independent mass parameters:
\be\label{GBmasses}
  \begin{aligned}
    M_{A(n)}^2 &= \frac{n^2}{R^2},\\[1ex]
    M_{Z(n)}^2 &= \frac{n^2}{R^2}+M^2_Z,\\[1ex]
    M_{W(n)}^2 &= \frac{n^2}{R^2}+M^2_W,\\[1ex]
    M_{H(n)}^2 &= \frac{n^2}{R^2}+2\mu^2.
  \end{aligned}
\ee

We introduce two sets of mass parameters appearing in the
fermion-scalar couplings. 
The mass parameters $m_x^{(i)}$ are
\be\label{mparameters}
  \begin{aligned}
    m_1^{(i)} &= \nor c\kkx in   + m_i s\kkx in,\\[1ex]
    m_2^{(i)} &= -\nor s\kkx in  + m_i c\kkx in,\\[1ex]
    m_3^{(i)} &= -M_W c\kkx in   + \nor \frac{m_i}{M_W} s\kkx
    in,\\[1ex]  
    m_4^{(i)} &= M_W s\kkx in    + \nor \frac{m_i}{M_W} c\kkx in,
  \end{aligned}
\ee
where $M_W$ and the {\it up-type} fermion masses $m_i$ on the r.h.s. are the
zero-mode masses. The parameters $c\kkx in$ and $s\kkx in$ denote the
cosine and sine respectively of the fermion mass mixing angle
$\alpha\kkx in$ defined in (\ref{fermionmassmixingangle}).
The mass parameters $M_x^{(i,j)}$ are
\be\label{Mparameters}
  \begin{aligned}
    M_1^{(i,j)}  &= m_j c_{i(n)},\\[1ex]
    M_2^{(i,j)}  &= m_j s_{i(n)},\\[1ex]
    M_3^{(i,j)}  &= \nor \frac{m_j}{M_W} c_{i(n)},\\[1ex]
    M_4^{(i,j)}  &= \nor \frac{m_j}{M_W} s_{i(n)},
  \end{aligned}
\ee
where again $M_W$ and the {\it down-type} fermion masses $m_j$ on the
r.h.s. are the zero-mode masses.

All momenta and fields are assumed to be incoming. 

\subsection{Propagators}
\label{FRpropagator}
The propagators are:\vspace{1ex}

for scalar fields $S=H,a^0,a^\pm,A_5,G^0,G^\pm$:\\[-1ex]

\begin{minipage}{40mm}
  \fbox{\includegraphics[]{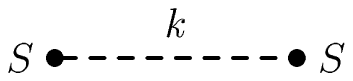}}
\end{minipage}
\begin{minipage}{12cm}
  $\displaystyle  = \frac{i}{k^2 - M^2 + i\epsilon},$
\end{minipage}\\[2ex]
with the masses
\be
  \renewcommand{\arraystretch}{1.2}
  \begin{array}{|c|cccccc|}\hline
    S & H\kk n & a^0\kk n & a^\pm\kk n & A\kkx 5n & G^0\kk n &
    G^\pm\kk n\\\hline  
    M & M\kkx Hn & M\kkx Zn & M\kkx Wn  & \sqrt\xi M\kkx An & \sqrt\xi
    M\kkx Zn & \sqrt\xi M\kkx Wn \\\hline
  \end{array}
\ee

\vspace{3ex}
for gauge bosons $V=A,Z,W^\pm$:\\[-1ex]

\begin{minipage}{43mm}
  \fbox{\includegraphics[]{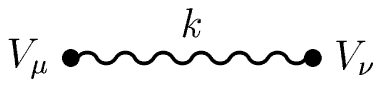}}
\end{minipage}
\begin{minipage}{12cm}
  $\displaystyle  =  \frac{-i}{k^2 -
    M^2 + i\epsilon}\left( g^{\mu\nu} -(1-\xi)\frac{k^\mu
      k^\nu}{k^2 -\xi M^2 + i\epsilon}\right),$
\end{minipage}\\[2ex]
with the masses
\be
  \renewcommand{\arraystretch}{1.2}
  \begin{array}{|c|ccc|}\hline
    V  &  A\kk n    &  Z\kk n    &  W^\pm\kk n  \\\hline
    M  &  M\kkx An  &  M\kkx Zn  &  M\kkx Wn    \\ \hline
  \end{array}
\ee
  
\vspace{3ex}
for fermion fields
$F=u,d,\mcQ_u,\mcQ_d,\mcU,\mcD,\nu,e,\mcL_\nu,\mcL_e,\mcE$:\\[-1ex]

\begin{minipage}{40mm}
  \fbox{\includegraphics[]{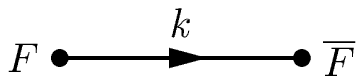}}
\end{minipage}
\begin{minipage}{12cm}
  $\displaystyle  = 
    \frac{i\left( \fslash k + m\right)}{k^2 - m^2 + i\epsilon},$
  \end{minipage}\\[2ex]
with the masses
\be
  \renewcommand{\arraystretch}{1.2}
  \begin{array}{|c|ccccccccccc|}\hline
    F & u & d & \mcQ\kkx un & \mcQ\kkx dn & \mcU\kk n & \mcD\kk n &
    \nu  & e & \mcL\kkx \nu n & \mcL\kkx en & \mcE\kk n \\\hline
    m & m_u & m_d & m\kkx un & m\kkx dn & m\kkx un & m\kkx dn & m_\nu
    & m_e & m\kkx \nu n & m\kkx en & m\kkx en \\\hline
  \end{array}
\ee  

\subsection{Vertices}
The Feynman rules for the vertices are:

\vspace{.5cm}
\begin{minipage}{38mm}
  \fbox{\includegraphics[]{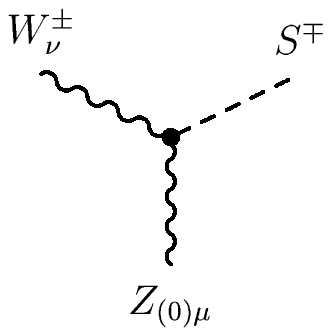}}
\end{minipage}
\begin{minipage}{12cm}
  $\displaystyle = \frac{g_2}{\cw M\kkx Wn} g_{\mu\nu} C.$
\end{minipage}

\begin{alignat}{2}
  &Z W^+ G^-  &\dpkt C &= - \sw^2 M_W^2 + \cw^2 \norsq,\\ 
  &Z W^- G^+  &\dpkt C &= \sw^2 M_W^2 - \cw^2 \norsq,\\
  &Z W^+ a^-  &\dpkt C &= -M_W \nor,\\
  &Z W^- a^+  &\dpkt C &= M_W \nor.  
\end{alignat}

\vspace{.5cm} 
\begin{minipage}{42mm}
  \fbox{\includegraphics[]{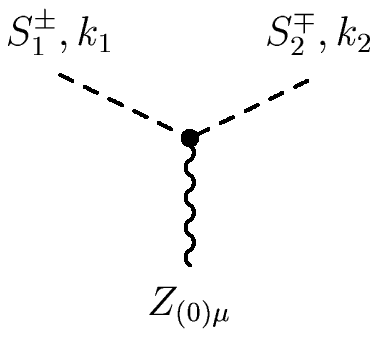}}
\end{minipage}
\begin{minipage}{12cm}
  $\displaystyle = \frac{i g_2}{2\cw M\kkx Wn^2 }(k_2-k_1)_\mu C.$
\end{minipage}

\begin{alignat}{2}
  &Z G^+ G^-  &\dpkt C &= -\left(\cw^2-\sw^2\right) M_W^2 - 2\cw^2
  \norsq,\\   
  &Z a^+ a^-  &\dpkt C &=  - 2\cw^2 M_W^2 -
  \left(\cw^2-\sw^2\right)\norsq,\\
  &Z G^+ a^-  &\dpkt C &= M_W \nor,\\ 
  &Z G^- a^+  &\dpkt C &= -M_W\nor.
\end{alignat}

\vspace{.5cm}
\begin{minipage}{44mm}
  \fbox{\includegraphics[]{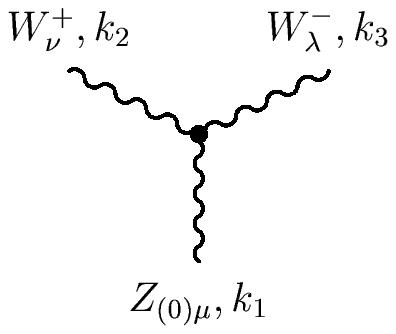}}
\end{minipage}
\begin{minipage}{12cm}
  $\displaystyle = ig_2\cw \bigl( g_{\mu\nu} (k_2
      -k_1)_\lambda + g_{\mu\lambda} (k_1 -k_3)_\nu +
    g_{\lambda\nu} (k_3 -k_2)_\mu \bigr).$ 
\end{minipage}
\vspace{-10ex}\be\ee\vspace{3ex}

\vspace{.5cm}
\begin{minipage}{36mm}
  \fbox{\includegraphics[]{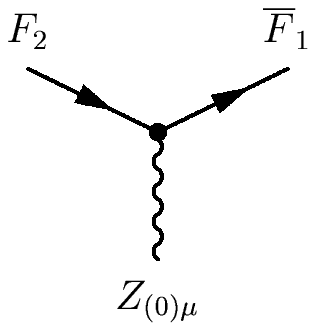}}
\end{minipage}
\begin{minipage}{12cm}
  $\displaystyle  = \frac{i g_2}{6 \cw} \gamma_\mu \left( P_L C_L +
    P_R C_R \right).$
\end{minipage}

\begin{alignat}{4}
  & Z \overline u_i u_i  &\dpkt 
  &\left\{\begin{array}{l}C_L = 3-4\sw^2,\\
      C_R = -4\sw^2,\end{array}\right.
  &\hspace{6ex}& Z \overline d_j d_j     &\dpkt 
  &\left\{\begin{array}{l}C_L = -3+2\sw^2,\\
      C_R = 2\sw^2,\end{array}\right.\\[1ex]
  &Z \overline \nu_i \nu_i     &\dpkt 
  &\left\{\begin{array}{l}C_L = 3,\\
      C_R = 0,\end{array}\right.
  &&Z \overline e_j e_j     &\dpkt 
  &\left\{\begin{array}{l}C_L = -3+6\sw^2,\\
      C_R = 6\sw^2 ,\end{array}\right.\\[1ex]
  & Z \overline \mcQ_i \mcQ_i     &\dpkt 
  &\left\{\begin{array}{l}C_L = -4\sw^2+3c\kkx in^2,\\
      C_R = -4\sw^2+3c\kkx in^2,\end{array}\right.
  &&Z \overline \mcU_i \mcU_i     &\dpkt 
  &\left\{\begin{array}{l}C_L = -4\sw^2+3s\kkx in^2,\\
      C_R = -4\sw^2+3s\kkx in^2,\end{array}\right.\\[1ex]
  & Z \overline \mcQ_i \mcU_i     &\dpkt 
  &\left\{\begin{array}{l}C_L = -3s\kkx in c\kkx in,\\
      C_R = 3s\kkx in c\kkx in,\end{array}\right.
  &&Z \overline \mcU_i \mcQ_i     &\dpkt 
  &\left\{\begin{array}{l}C_L = -3s\kkx in c\kkx in,\\
      C_R = 3s\kkx in c\kkx in.\end{array}\right.
\end{alignat}

\vspace{.5cm}
\begin{minipage}{36mm}
  \fbox{\includegraphics[]{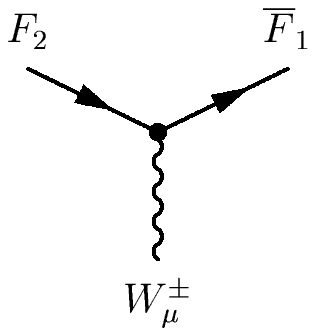}}
\end{minipage}
\begin{minipage}{12cm}
  $\displaystyle  = \frac{i g_2}{\sqrt{2}} \gamma_\mu P_L C_L.$
\end{minipage}

\begin{alignat}{4}
  & W^+~\overline u_i d_j       &\dpkt C_L &= V_{ij},
  &\hspace{12ex}&W^-~\overline d_j u_i     &\dpkt C_L &= V_{ij}^*,\\
  & W^+~\overline \mcQ_i d_j    &\dpkt C_L &= c_{i(n)} V_{ij},
  &&W^-~\overline d_j\mcQ_i     &\dpkt C_L &= c_{i(n)} V_{ij}^*,\\
  & W^+~\overline \mcU_i d_j    &\dpkt C_L &= -s_{i(n)} V_{ij},
  &&W^-~\overline d_j\mcU_i     &\dpkt C_L &= -s_{i(n)} V_{ij}^*,\\
  & W^+~\overline \nu _i e_j    &\dpkt C_L &= \delta_{ij},
  &&W^-~\overline e_j \nu_i     &\dpkt C_L &= \delta_{ij},\\
  & W^+~\overline \nu_i \mcL_j  &\dpkt C_L &= \delta_{ij},
  &&W^-~\overline \mcL_j \nu_i  &\dpkt C_L &= \delta_{ij},\\
  & W^+~\overline \nu_i \mcE_j  &\dpkt C_L &= 0,
  &&W^-~\overline \mcE_j \nu_i  &\dpkt C_L &= 0.
\end{alignat}

\vspace{.5cm}
\begin{minipage}{36mm}
  \fbox{\includegraphics[]{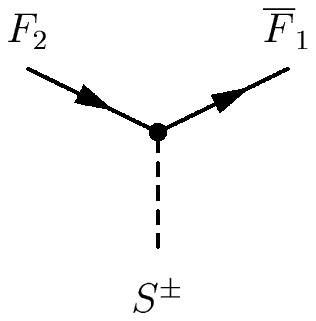}}
\end{minipage}
\begin{minipage}{12cm}
  $\displaystyle  = \frac{g_2}{\sqrt{2} M_{W(n)}} (P_L
  C_L + P_R C_R).$
\end{minipage}

\begin{alignat}{4}
  & G^+~\overline u_i d_j   &\dpkt
  &\left\{\begin{array}{l}C_L = -m_i V_{ij},\\
      C_R = m_j V_{ij},\end{array}\right.
  &\hspace{8ex}&G^-~\overline d_j u_i     &\dpkt
  &\left\{\begin{array}{l}C_L = -m_j V_{ij}^*,\\
      C_R = m_i V_{ij}^*,\end{array}\right.\\[1ex]
  & G^+~\overline\mcQ_i d_j   &\dpkt
  &\left\{\begin{array}{l}C_L = -m_1^{(i)} V_{ij},\\
      C_R = M_1^{(i,j)} V_{ij},\end{array}\right.
  &&G^-~\overline d_j\mcQ_i   &\dpkt
  &\left\{\begin{array}{l}C_L = -M_1^{(i,j)} V_{ij}^*,\\
      C_R = m_1^{(i)} V_{ij}^*,\end{array}\right.\\[1ex]
  & G^+~\overline\mcU_i d_j     &\dpkt
  &\left\{\begin{array}{l}C_L = m_2^{(i)} V_{ij},\\
      C_R = -M_2^{(i,j)} V_{ij},\end{array}\right.
  &&G^-~\overline d_j\mcU_i     &\dpkt
  &\left\{\begin{array}{l}C_L = M_2^{(i,j)} V_{ij}^*,\\
      C_R = -m_2^{(i)} V_{ij}^*,\end{array}\right.\\[1ex]
  & G^+~\overline \nu_i e_j   &\dpkt
  &\left\{\begin{array}{l}C_L = 0,\\
      C_R = m_j \delta_{ij},\end{array}\right.
  &&G^-~\overline e_j \nu_i     &\dpkt
  &\left\{\begin{array}{l}C_L = -m_j \delta_{ij},\\
      C_R = 0,\end{array}\right.\\[1ex]
  & G^+~\overline \nu_i \mcL_j   &\dpkt
  &\left\{\begin{array}{l}C_L = 0,\\
      C_R = m_1^{(j)} \delta_{ij},\end{array}\right.
  &&G^-~\overline \mcL_j \nu_i   &\dpkt
  &\left\{\begin{array}{l}C_L = -m_1^{(j)} \delta_{ij},\\
      C_R = 0,\end{array}\right.\\[1ex]
  & G^+~\overline \nu_i \mcE_j     &\dpkt
  &\left\{\begin{array}{l}C_L = 0,\\
      C_R = -m_2^{(j)} \delta_{ij},\end{array}\right.
  &&G^-~\overline \mcE_j\nu_i     &\dpkt
  &\left\{\begin{array}{l}C_L = m_2^{(j)} \delta_{ij},\\
      C_R = 0,\end{array}\right.\\[1ex]
  & a^+~\overline\mcQ_i d_j   &\dpkt
  &\left\{\begin{array}{l}C_L = -m_3^{(i)} V_{ij},\\
      C_R = M_3^{(i,j)} V_{ij},\end{array}\right.
  &\hspace{8ex}&a^-~\overline d_j\mcQ_i &\dpkt
  &\left\{\begin{array}{l}C_L = -M_3^{(i,j)} V_{ij}^*,\\
      C_R = m_3^{(i)} V_{ij}^*,\end{array}\right.\\[1ex]
  & a^+~\overline\mcU_i d_j     &\dpkt
  &\left\{\begin{array}{l}C_L = m_4^{(i)} V_{ij},\\
      C_R = -M_4^{(i,j)} V_{ij},\end{array}\right.
  &&a^-~\overline d_j\mcU_i     &\dpkt
  &\left\{\begin{array}{l}C_L = M_4^{(i,j)} V_{ij}^*,\\
      C_R = -m_4^{(i)} V_{ij}^*,\end{array}\right.\\
  & a^+~\overline \nu_i \mcL_j   &\dpkt
  &\left\{\begin{array}{l}C_L = 0,\\
      C_R = m_3^{(j)} \delta_{ij},\end{array}\right.
  &&a^-~\overline \mcL_j \nu_i &\dpkt
  &\left\{\begin{array}{l}C_L = -m_3^{(j)} \delta_{ij},\\
      C_R = 0,\end{array}\right.\\[1ex]
  & a^+~\overline \nu_i \mcE_j     &\dpkt
  &\left\{\begin{array}{l}C_L = 0,\\
      C_R = -m_4^{(j)} \delta_{ij},\end{array}\right.
  &&a^-~\overline \mcE_j \nu_i     &\dpkt
  &\left\{\begin{array}{l}C_L = m_4^{(j)} \delta_{ij},\\
      C_R = 0.\end{array}\right.
\end{alignat}

\newsection{Different Contributions to \boldmath{$\Delta F=2$} Box Diagrams}
The contributions from different sets of diagrams to the functions
$F(x_{t(n)},x_{u(n)})$ in  (\ref{SN}) are given as follows

  \bea
    F_{WW(n)} = \frac{M_W^2}{M_{W(n)}^2}~ U(x_{t(n)}, x_{u(n)}),
  \eea
  \bea
    \lefteqn{
    F_{WG(n)} = -2\frac{M_W^2 m_{t(n)}m_{u(n)}}{M_{W(n)}^6} \left[
      m_1^{(t)} c_{t(n)} + m_2^{(t)} s_{t(n)} \right]
      \left[m_1^{(u)} c_{u(n)} + m_2^{(u)} s_{u(n)} \right]}\nnl[2mm]
      & &{}\times\tilde U(x_{t(n)}, x_{u(n)}), 
  \eea
  \bea
    \lefteqn{
    F_{Wa(n)} = -2\frac{M_W^2 m_{t(n)}m_{u(n)}}{M_{W(n)}^6} \left[
      m_3^{(t)} c_{t(n)} + m_4^{(t)} s_{t(n)} \right]
      \left[m_3^{(u)} c_{u(n)} + m_4^{(u)} s_{u(n)} \right]}\nnl[2mm]
      & &{}\times\tilde U(x_{t(n)}, x_{u(n)}), 
  \eea
  \bea
    \lefteqn{
    F_{Ga(n)} = \frac12\frac{M_W^2}{M_{W(n)}^6}
    \left[m_1^{(t)} m_3^{(t)} + m_2^{(t)}
      m_4^{(t)}\right]
    \left[m_1^{(u)} m_3^{(u)} + m_2^{(u)} m_4^{(u)} \right]}\nnl[2mm]
    & &{}\times U(x_{t(n)}, x_{u(n)}),
  \eea
  \bea
    F_{GG(n)} = \frac14\frac{M_W^2}{M_{W(n)}^6}
    \left[(m_1^{(t)})^2 + (m_2^{(t)})^2 \right]\left[(m_1^{(u)})^2 +
      (m_2^{(u)})^2 \right] U(x_{t(n)}, x_{u(n)}),
  \eea
  \bea
    F_{aa(n)} = \frac14\frac{M_W^2}{M_{W(n)}^6}
    \left[(m_3^{(t)})^2 + (m_4^{(t)})^2\right]\left[(m_3^{(u)})^2 +
      (m_4^{(u)})^2 \right] U(x_{t(n)}, x_{u(n)}).
  \eea

  The functions $U$ and $\tilde U$ are defined as follows
  \bea\label{U1}
    U (x_t, x_u) = \frac{x_t^2\log{x_t}}{(x_t-x_u)(1-x_t)^2} +
    \frac{x_u^2\log{x_u}}{(x_u-x_t)(1-x_u)^2} + \frac{1}{(1-x_u)(1-x_t)}, 
  \eea
  \bea\label{U2}
    U (x_t, x_t) = \frac{2 x_t\log{x_t}}{(1-x_t)^3} +
    \frac{1+x_t}{(1-x_t)^2}, 
  \eea
  \bea\label{U3}
    \tilde U (x_t, x_u) = \frac{x_t\log{x_t}}{(x_t-x_u)(1-x_t)^2} +
    \frac{x_u\log{x_u}}{(x_u-x_t)(1-x_u)^2} + \frac{1}{(1-x_u)(1-x_t)}, 
  \eea
  \bea\label{U4}
    \tilde U (x_t, x_t) = \frac{(1+x_t)\log{x_t}}{(1-x_t)^3} +
    \frac{2}{(1-x_t)^2}.
  \eea

\begin{figure}[hbt] \centering
  \subfigure[]{
        \includegraphics[]{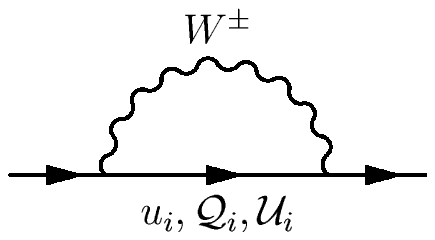} }\hspace{0.4cm}
   \subfigure[]{
        \includegraphics[]{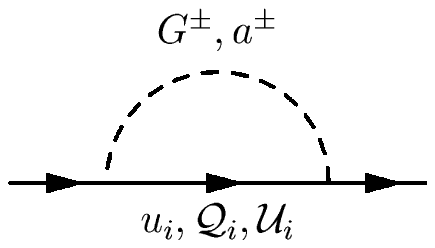} }
 
    \caption[]{\small\label{selfenergydiagrams} Self-energy diagrams
    neccessary for calculating the electroweak counter term as
    discussed in \cite{BB1}.}
  \end{figure}

\newsection{Different Contributions to \boldmath{$Z^0$}-Penguin Diagrams}
\label{apppenguin}
  The contributions of the diagrams in fig. \ref{penguindiagrams}
  to the  functions $F(x_f)$ in (\ref{CN}) are given as follows 
  \begin{eqnarray}
  F_1(x_{f(n)}) &=&\frac{1}{8} \left(c_{f(n)} ^4 + 
  s_{f(n)}^4 - \frac{4}{3} \sw^2\right) \nnl && \times \left[ \Delta + 
  \ln  \frac{\mu^2}{m^2_{f(n)}} +  h_q\left(x_{f(n)}\right) 
-\frac{3}{2} 
  -   2 x_{f(n)} h_q\left(x_{f(n)}\right)\right],\\
  F_2(x_{f(n)}) &=  & \frac{1}{4} c_{f(n)}^2  s_{f(n)}^2 \left[ \Delta + 
  \ln\frac{\mu^2}{m_{f(n)}^2} -\frac{3}{2} 
  + h_q\left(x_{f(n)}\right)
  + 2 x_{f(n)} h_q\left(x_{f(n)}\right){} \right],\\
  F_3(x_{f(n)}) &=  &\frac{\displaystyle 1}{\displaystyle 16  m^2_{W(n)} }
  \left[ \left( (m^{{(f)}}_{1})^2+ (m^{{(f)}}_{3})^2\right)\left(c_{f{(n)}}^{2} -\frac{4}{3} \sw^2 \right)
    +\left((m^{{(f)}}_{2})^2+ (m^{{(f)}}_{4})^2\right)  \left(s_{f(n)}^2-\frac{4}{3} \sw^2 \right)\right]\nnl
  &&\times \left[ \Delta + 
      \ln \frac{\mu^2}{m_{f(n)}^2}-\frac{1}{2}  +  h_q\left(x_{f(n)}\right) 
      - 2 x_{f(n)} h_q\left(x_{f(n)}\right)\right],\\
  F_4(x_{f(n)})&=& -\frac{1}{8 M^2_{W(n)}} 
      \left( m^{{(f)}}_{1}\, m^{{(f)}}_{2}
    + m^{{(f)}}_{3}\, m^{{(f)}}_{4}\right)~c_{f(n)}~s_{f(n)} 
  \nnl &&\times \left[ \Delta + 
  \ln\frac{\mu^2}{m_{f(n)}^2}-\frac{1}{2}  +  h_q\left(x_{f(n)}\right) 
    + 2 x_{f(n)} h_q\left(x_{f(n)}\right)\right],\\
  F_5(x_{f(n)})&=& -\frac{3}{4} \,\cw^2 \,\left[ \Delta + 
  \ln \frac{\mu^2}{M_{W(n)}^2}-\frac{1}{6}  - x_{f(n)} h_w\left(x_{f(n)}\right)\right],\\
  F_6(x_{f(n)}) &= &-\frac{1}{16\, m^{4}_{W(n)}} 
  \left[ \left( \left( 1-2\,\sw^2 \right) M_W^2 + 2 \cw^2 \norsq \right) 
     \left((m^{{(f)}}_{1})^2+ (m^{{(f)}}_{2})^2\right)  \right. \nnl
    &&\left.+ \left(\left(1-2\,\sw^2 \right) \norsq   + 2 \cw^2 M_W^2 \right)  \left((m^{{(f)}}_{3})^2+ (m^{{(f)}}_{4})^2\right)
        \right]\nnl
  &&\times \left[ \Delta + 
    \ln \frac{\mu^2}{M_{W(n)}^2}+\frac{1}{2}  - x_{f(n)} h_w\left(x_{f(n)}\right) \right] ,\\
  F_7(x_{f(n)}) &=&\frac{M_W  }{8 M^{4}_{W{(n)}}} \frac{n}{R}
  \left( m^{{(f)}}_{1} m^{{(f)}}_{3} +  m^{{(f)}}_{2}
  m^{{(f)}}_{4}\right)\nnl && \times
  \left[ \Delta + 
    \ln \frac{\mu^2}{M_{W(n)}^2}+\frac{1}{2}  - x_{f(n)}\, h_w\left(x_{f(n)}\right)\right] ,\\
  F_8(x_{f(n)}) &=  &  \,\frac{m_{f(n)} }{2 M^{4}_{W{(n)}}}
  \left[ \left(\sw^2 M_W^2 - c_w^2 \norsq\right) \,\left(m^{{(f)}}_{1} ~c_{f(n)} + m^{{(f)}}_{2} ~s_{f(n)} \right)\right. \nnl && \left. + M_W
       {n\over R } \,\left( m^{{(f)}}_{3} ~c_{f(n)} 
       + m^{{(f)}}_{4} ~s_{f(n)} \right)\right]\,
       h_w\left(x_{f(n)}\right).
   \end{eqnarray}
  Here 
 \be
 \Delta=\frac{1}{\epsilon}+\ln 4\pi-\gamma_E, \qquad D=4-2\epsilon
 \ee
 and
 the functions $h_q$ and $h_w$ are given by:
  \begin{eqnarray}
    h_q(x) &=& \frac{1}{1-x} + \frac{\ln x}{(1-x)^2}, \\
    h_w(x) &=& \frac{1}{1-x} + \frac{x \ln x}{(1-x)^2} ~.
  \end{eqnarray}
  Finally, the contributions  from counter terms corresponding to 
  the self-energy diagrams of
  fig. \ref{selfenergydiagrams} that should be added to the functions 
  $F_i$ are given by
  \begin{eqnarray}
     \Delta S_1\left(x_{f(n)} \right) \!\!\! &=&  \!\!\!\frac14 \left(\Delta
    -\frac{1}{2} \left( \frac{1+x_{f(n)}}{1-x_{f(n)}} +  \frac{2
    x_{f(n)}^2 \ln x_{f(n)} }{(1-x_{f(n)})^2} \right) - 
    \ln { \frac{M^2_{W(n)}}{\mu^2}} \right),\\
     \Delta S_2\left(x_{f(n)} \right)\! \!\! &=& \! \!\!\frac18 \left(
    1+ x_f \right)  \left(\Delta +\frac{1}{2} 
\left( \frac{1-3x_{f(n)}}{1-x_{f(n)}} -
          \frac{2 x_{f(n)}^2\ln x_{f(n)} }{(1-x_{f(n)})^2} \right)   - 
    \ln { \frac{M^2_{W(n)}}{\mu^2}} \right) \!.
  \end{eqnarray}

\newsection{Closed form for the $\Delta C(x_t,1/R)$ function}
\label{CSUMAPP}
We express the logarithms of~(\ref{CFIN}) as integrals $I_1$ and $I_2$
 \begin{eqnarray}
  x_n \ln \frac{x_n + x_t}{1 + x_n} &=& I_1(x_t) - I_1(1) +x_t -1,\\
  \ln \frac{x_n + x_t}{1 + x_n} &=& I_2(x_t) - I_2(1)
\end{eqnarray}
with
\begin{eqnarray}
  I_1(a) &=& a^2 \int_0^1 dy \frac{y}{a^2 y + x_n},\\
  I_2(a) &=& a \int_0^1 dy \frac{1}{a \,y + x_n}.
\end{eqnarray}
Next, we interchange integration and summation. The integrands can now be summed using the relation
\begin{eqnarray}
	\sum_{n=1}^{\infty} \frac{b}{c + n^2} =\frac{b\,\left( {\sqrt{c}}\,\pi \,\coth ({\sqrt{c}}\,\pi ) -1 \right) }{2\,c}.
\end{eqnarray}
This allows us to derive a closed form for the sum
\begin{eqnarray}\label{cclosedfin}
	\sum_{n=1}^{\infty} C_n ( x_t,x_n) &=&  \frac{
  \left(  x_t -7 \right)  x_t  }{16 \left( 1 - x_t \right) } -\frac{m_W \pi R x_t}{16 {\left( 1 - x_t \right) }^2}
 \nonumber\\&& \times
\int_0^1 dy \Bigg[
  \frac{3 \left( 1 + x_t \right)  }{\sqrt{y}}  
     \left( \coth (m_W \pi  R {\sqrt{y}}) -
       x_t^\frac12 \coth (m_t \pi  R {\sqrt{ y}}) \right) 
\nonumber\\&&  + \left( x_t -7\right) {\sqrt{y}} 
     \left( \coth (m_W \pi  R {\sqrt{y}}) - x_t^{\frac{3}{2}} 
        \coth (m_t \pi  R {\sqrt{ y}}) \right) 
\Bigg].
\end{eqnarray} 
Expanding in $(m_W \pi R\sqrt{y}, m_t \pi R \sqrt{y})$ and integrating~(\ref{cclosedfin}) leads to the expression in~(\ref{csummed}).

\newsection{Different Contributions to \boldmath{$\Delta F=1$} Box
  Diagrams} 

  The one loop amplitude of the KK excitations in the diagrams in
  fig. \ref{boxmumu} is a sum of
  contributions coming from the various bosonic fields in the loop
    \begin{eqnarray}\label{sixG}
    G_n(x_{f(n)},x_{e(n)}) \!& =& \! G_{WW(n)} + G_{WG(n)} + G_{Wa(n)} +
    G_{Ga(n)} + G_{GG(n)} + G_{aa(n)},\\\label{sixH}
    H_n(x_{f(n)},x_{\nu(n)}) \! &= & \!H_{WW(n)} + H_{WG(n)} + H_{Wa(n)} +
    H_{Ga(n)} + H_{GG(n)} + H_{aa(n)}.
  \end{eqnarray}
  The explicit results for $B^{\nu\bar\nu}_n
  (x_t,x_n)$ and $ B^{\mu\bar\mu}_n (x_t,x_n)$ are given by
  \begin{eqnarray}\label{BN}
       B^{\nu\bar\nu}_n (x_t,x_n) &=& G_n(x_{t(n)},x_{e(n)}) -
  G_n(x_{u(n)},x_{e(n)}),\nnl
  B^{\mu\bar\mu}_n (x_t,x_n) &=& H_n(x_{t(n)},x_{\nu(n)}) -
  H_n(x_{u(n)},x_{\nu(n)}),
  \end{eqnarray}
  where the results for the $ \nu
  \bar\nu$-box diagrams take the form
  \begin{eqnarray}
    G_{WW(n)} &=& -\frac{M_W^2}{M_{W(n)}^2}~ U(x_{f(n)}, x_{e(n)}),\\
    G_{WG(n)} &=& \frac12 \frac{M_W^2 m_{f(n)}m_{e(n)}}{M_{W(n)}^6} \left[
      m_1^{(f)} c_{f(n)} + m_2^{(f)} s_{f(n)} \right] ~m_1^{(e)}~
      \tilde U(x_{f(n)}, x_{e(n)}), \\
       G_{Wa(n)} &=& \frac12\frac{M_W^2 m_{f(n)}m_{e(n)}}{M_{W(n)}^6} \left[
      m_3^{(f)} c_{f(n)} + m_4^{(f)} s_{f(n)} \right]~ m_3^{(e)} ~
      \tilde U(x_{f(n)}, x_{e(n)}), \\
      G_{Ga(n)} &=& -\frac{1}{8}\frac{M_W^2}{M_{W(n)}^6}
    \left[m_1^{(f)} m_3^{(f)} + m_2^{(f)}
      m_4^{(f)}\right]~m_1^{(e)}\, m_3^{(e)}~ U(x_{f(n)}, x_{e(n)}),\\
    G_{GG(n)} &=& -\frac{1}{16}\frac{M_W^2}{M_{W(n)}^6}
    \left[(m_1^{(f)})^2 + (m_2^{(f)})^2 \right] (m_1^{(e)})^2~  U(x_{f(n)}, x_{e(n)}),\\
    G_{aa(n)} &=& -\frac{1}{16}\frac{M_W^2}{M_{W(n)}^6}
    \left[(m_3^{(f)})^2 + (m_4^{(f)})^2\right] (m_3^{(e)})^2 ~  U(x_{f(n)}, x_{e(n)}),
  \end{eqnarray}
  and for the $\mu\bar\mu$-box diagrams we get 
  \begin{eqnarray}
    H_{WW(n)} &=&-\frac14 \frac{M_W^2}{M_{W(n)}^2}~ U(x_{f(n)}, x_{\nu(n)}),\\
    H_{WG(n)} &=& \frac12\,\frac{M_W^2 m_{f(n)}m_{\nu(n)}}{M_{W(n)}^6} \left[
      m_1^{(f)} c_{f(n)} + m_2^{(f)} s_{f(n)} \right] ~m_1^{(\nu)}~
      \tilde U(x_{f(n)}, x_{\nu(n)}), \\
       H_{Wa(n)} &=& \frac12\,\frac{M_W^2 m_{f(n)}m_{\nu(n)}}{M_{W(n)}^6} \left[
      m_3^{(f)} c_{f(n)} + m_4^{(f)} s_{f(n)} \right]~ m_3^{(\nu)} ~
      \tilde U(x_{f(n)}, x_{\nu(n)}), \\
      H_{Ga(n)} &=& -\frac{1}{8} \frac{M_W^2}{M_{W(n)}^6}
    \left[m_1^{(t)} m_3^{(f)} + m_2^{(f)}
      m_4^{(f)}\right]~m_1^{(\nu)}\, m_3^{(\nu)}~ U(x_{f(n)}, x_{\nu(n)}),\\
     H_{GG(n)} &=&-\frac{1}{16} \frac{M_W^2}{M_{W(n)}^6}
   \left[(m_1^{(f)})^2 + (m_2^{(f)})^2 \right] (m_1^{(\nu)})^2~  U(x_{f(n)}, x_{\nu(n)}),\\
    H_{aa(n)} &=& -\frac{1}{16}\frac{M_W^2}{M_{W(n)}^6}
   \left[(m_3^{(f)})^2 + (m_4^{(f)})^2\right] (m_3^{(\nu)})^2 ~  
U(x_{f(n)}, x_{\nu(n)}).
 \end{eqnarray}
The functions $U$ and $\tilde U$ are defined in (\ref{U1})--(\ref{U4}). 
 We have taken into account the overall minus sign in (\ref{hyll}).

  Summing all the contributions we find 
  \begin{eqnarray}
   B^{\nu\bar\nu}_n (x_t,x_n)\!\! &=&\!\!  \frac{17 x_t +  18 x_n  
     x_t -9 x_n }
     {16 \left( x_t -1\right) }
   + \frac{x_n 
     \left( 26 x_t + 
       9 x_n 
         + 18 x_n x_t \right)}
 {16 x_t}
         \ln \frac{x_n}
       {1 + x_n}  \nonumber\\
  &&-\frac{\left( x_n + 
       x_t \right)  
     \left( 9 x_n + 
       17 x_t \right)}
{16 {\left( x_t -1 \right)
         }^2 x_t}  
     \ln \frac{x_n + 
         x_t}{1 + x_n}  ~,
   \end{eqnarray}

  \begin{eqnarray}
  B^{\mu\bar\mu}_n (x_t,x_n)\!\! &=&\!\!  \frac{3 x_n + 5 x_t - 
     6 x_n x_t}{16 (x_t -1)} - 
  \frac{x_n 
     \left( 3 x_n + 6 x_n x_t - 2 x_t \right)}{16 x_t} 
         \ln \frac{x_n}{1+ x_n} \nonumber\\&&+ 
  \frac{\left( 3 x_n - 
       5 x_t \right)  
     \left( x_n + 
       x_t \right)}{16 {\left( x_t -1 \right)
         }^2 x_t}  
     \ln \frac{x_n + x_t}
         {1 + x_n} ~.
   \end{eqnarray}

\newpage

\newcommand{\np}[3]{Nucl.~Phys. {\bf B#1} (#2) #3}
\newcommand{\pl}[3]{Phys.~Lett. {\bf B#1} (#2) #3}
\newcommand{\pr}[3]{Phys.~Rev.  {\bf D#1} (#2) #3}
\newcommand{\prl}[3]{Phys.~Rev. Lett. {\bf #1} (#2) #3}
\newcommand{\prp}[3]{Phys.~Rept. {\bf #1} (#2) #3}
\newcommand{\zpc}[3]{Z.~Phys. {\bf C#1} (#2) #3}
\newcommand{\hep}[2]{[arXiv:hep-#1/#2]}

\renewcommand{\baselinestretch}{0.95}

\vfill\eject


\begin{thebibliography}{99}
    \bibitem{appelquist:01}T.~Appelquist, H.-C.~Cheng and B.~A.~Dobrescu,
      \pr{64}{2001}{035002}, \hep{ph}{0012100}.
\bibitem{AGDEWU}
K.~Agashe, N.~G.~Deshpande and G.~H.~Wu,
Phys.\ Lett.\ B {\bf 514} (2001) 309,
[arXiv:hep-ph/0105084].
\bibitem{APDO} 
K.~Agashe, N.~G.~Deshpande and G.~H.~Wu,
Phys.\ Lett.\ B {\bf 511} (2001) 85,
[arXiv:hep-ph/0103235];
T.~Appelquist and B.~A.~Dobrescu,
Phys.\ Lett.\ B {\bf 516} (2001) 85,
[arXiv:hep-ph/0106140].
\bibitem{COLL0}
C. Macesanu, C.D. McMullen and S. Nandi, 
Phys.\ Rev.\ D {\bf 66} (2002) 015009
[arXiv:hep-ph/0201300].
\bibitem{COLL1}
H.~C.~Cheng, K.~T.~Matchev and M.~Schmaltz,
Phys.\ Rev.\ D {\bf 66} (2002) 056006
[arXiv:hep-ph/0205314].
\bibitem{COLL2}
T.~G.~Rizzo,
Phys.\ Rev.\ D {\bf 64} (2001) 095010
[arXiv:hep-ph/0106336].
\bibitem{COLL3}
F.~J.~Petriello,
JHEP {\bf 0205} (2002) 003
[arXiv:hep-ph/0204067].
\bibitem{SETA} 
G.~Servant and T.~M.~Tait,
[arXiv:hep-ph/0206071].
\bibitem{DARK}
H.~C.~Cheng, J.~L.~Feng and K.~T.~Matchev,
[arXiv:hep-ph/0207125]; 
D.~Hooper and G.~D.~Kribs,
[arXiv:hep-ph/0208261];
G.~Servant and T.~M.~Tait,
[arXiv:hep-ph/0209262];
D.~Majumdar,
[arXiv:hep-ph/0209277];
G.~Bertone, G.~Servant and G.~Sigl,
[arXiv:hep-ph/0211342].
\bibitem{APYE} T. Appelquist and H. Yee, [arXiv:hep-ph/0211023].
\bibitem{BPSW02}
A.J. Buras, A. Poschenrieder, M. Spranger and A. Weiler, in preparation. 
\bibitem{GIM}
{ S.L. Glashow, J. Iliopoulos and L. Maiani}
{ Phys. Rev.} {\bf D 2} (1970) 1285.
\bibitem{OLPASA}
J. Papavassiliou and A. Santamaria, { Phys. Rev.} {\bf D 63} (2001) 016002.
J.F. Oliver, J. Papavassiliou and A. Santamaria, [arXiv:hep-ph/0209021].
\bibitem{UUT}
A.J. Buras, P. Gambino, M. Gorbahn, S. J\"ager and L. Silvestrini,
{ Phys. Lett.} {\bf B500} (2001) 161.
\bibitem{IL}
{ T. Inami and C.S. Lim,}
{ Progr. Theor. Phys.} {\bf 65} (1981) 297.
\bibitem{BSS}
{ A.J. Buras, W. Slominski and H. Steger,} { Nucl. Phys.} {\bf B238} 
(1984) 529, { Nucl. Phys.} {\bf B245} (1984) 369.
\bibitem{PBE0}
{ G. Buchalla, A.J. Buras and M.K. Harlander,} { Nucl. Phys.}
 {\bf B 349} (1991) 1.
\bibitem{CHHUKU}
D. Chakraverty, K. Huitu and A. Kundu,
[arXiv:hep-ph/0212047].
\bibitem{rueckl}
A.~Muck, A.~Pilaftsis and R.~Ruckl,
[arXiv:hep-ph/0210410];
A.~Muck, A.~Pilaftsis and R.~Ruckl,
[arXiv:hep-ph/0209371];
A.~Muck, A.~Pilaftsis and R.~Ruckl,
[arXiv:hep-ph/0203032];
A.~Muck, A.~Pilaftsis and R.~Ruckl,
Phys.\ Rev.\ D {\bf 65} (2002) 085037,
[arXiv:hep-ph/0110391].
\bibitem{Georgi}
H.~Georgi, A.~K.~Grant and G.~Hailu,
Phys.\ Rev.\ D {\bf 63} (2001) 064027,
[arXiv:hep-ph/0007350].
\bibitem{Giedt}
J.~Giedt,
[arXiv:hep-ph/0204315].
\bibitem{Cheng:2002iz}
H.~C.~Cheng, K.~T.~Matchev and M.~Schmaltz,
Phys.\ Rev.\ D {\bf 66} (2002) 036005
[arXiv:hep-ph/0204342].
\bibitem{BJW90}
{ A.J. Buras, M. Jamin, and P.H. Weisz,}
{ Nucl. Phys.} {\bf B347} (1990) 491.
\bibitem{UKJS}
J. Urban, F. Krauss, U. Jentschura and G. Soff, 
{ Nucl. Phys.} {\bf B523} (1998) 40.
\bibitem{Erice}
A.J. Buras, lectures at the International Erice School, 
August, 2000, [arXiv:hep-ph/0101336].
\bibitem{DIDUGH}
K.R. Dienes, E. Dudas and T. Gherghetta, { Nucl. Phys.} {\bf B537} (1999) 47. 
\bibitem{HNa}
{ S. Herrlich and U. Nierste,}
{ Nucl. Phys.} {\bf B419} (1994) 292 and U. Nierste, recent update. 
 \bibitem{HNb}
{ S.~Herrlich and  U.~Nierste},
{ Phys. Rev.} {\bf D52} (1995) 6505; 
{ Nucl. Phys.} {\bf B476} (1996) 27.
\bibitem{WO}
{ L. Wolfenstein}, { Phys. Rev. Lett.} {\bf 51} (1983) 1945.
\bibitem{BLO}
{ A.J. Buras, M.E. Lautenbacher and G. Ostermaier,}
{ Phys. Rev.} {\bf D 50} (1994) 3433.
\bibitem{ref:lepbosc} 
               LEP Working group on oscillations :\\ 
       http://lepbosc.web.cern.ch/LEPBOSC/combined\_results/amsterdam\_2002/
\bibitem{lellouch}  
 L. Lellouch, [arXiv:hep-ph/0211359]; D. Becirevic, [arXiv:hep-ph/0211340]. 
\bibitem{Jamin} A.A. Penin and M. Steinhauser, 
{Phys.\ Rev.} {\bf D65} (2002) 054006; M. Jamin and B.O. Lange, 
{Phys.\ Rev.} {\bf D65} (2002) 056005; K. Hagiwara, S. Narison and D. Nomura,~[arXiv:hep-ph/0205092].
\bibitem{BUPAST} A.J. Buras, F. Parodi and A. Stocchi,~[arXiv:hep-ph/0207101].
\bibitem{BaBar}
B. Aubert et al., BaBar Collaboration, [arXiv:hep-ex/0207042]. 
\bibitem{Belle}
K. Abe et al., Belle Collaboration, [arXiv:hep-ex/0208025].
\bibitem{NIR02} Y. Nir, [arXiv:hep-ph/0110278].
\bibitem{DAGIISST}
G. D`Ambrosio, G.F. Giudice, G. Isidori and A. Strumia, [arXiv:hep-ph/0207036].
\bibitem{C00}
M. Ciuchini, G. D'Agostini, E. Franco, V. Lubicz, G. Martinelli, 
F. Parodi, P. Roudeau, A. Stocchi,
JHEP 0107(2001) 013,~[arXiv:hep-ph/0012308].
\bibitem{achille}  
A. Stocchi, [arXiv:hep-ph/0211245]. 
\bibitem{AJB02} A.J. Buras,~[arXiv:hep-ph/0210291] and references therein.
\bibitem{Lacker}
A. H\"ocker, H. Lacker, S. Laplace and F. Le Diberder, 
Eur. Phys. J. {\bf C21} (2001) 225 and http://ckmfitter.in2p3.fr.
\bibitem{PAST}
Provided by F. Parodi and A. Stocchi.
\bibitem{BB4}
{ G. Buchalla and A.J. Buras}, 
{ Phys. Lett.} {\bf B333} (1994) 221, Phys. Rev. {\bf D54} (1996) 6782. 
\bibitem{GRNI}
Y. Grossman and Y. Nir, { Phys. Lett.} {\bf B398} (1997) 163.
\bibitem{BB2}
{ G. Buchalla and A.J. Buras,}
{ Nucl. Phys.} {\bf B 400} (1993) 225.
\bibitem{MU98}
M. Misiak and J. Urban, { Phys. Lett.} {\bf B541} (1999) 161.
\bibitem{BB98}
G. Buchalla and A.J. Buras, { Nucl. Phys.} {\bf B 548} (1999) 309.
\bibitem{BB3}
{ G. Buchalla and A.J. Buras,}
{ Nucl. Phys.} {\bf B 412} (1994) 106.
\bibitem{BB1}
{ G. Buchalla and A.J. Buras,}
{ Nucl. Phys.} {\bf B 398} (1993) 285.
\bibitem{oliver:2002}
J.~F.~Oliver, J.~Papavassiliou and A.~Santamaria,
arXiv:hep-ph/0212391.
\bibitem{PDG}
K. Hagiwara et al., ``Review of Particle Physics", 
{ Phys. Rev.} {\bf D 66} (2002) 010001.
\bibitem{MP}
{ W. Marciano and Z. Parsa}, Phys. Rev. {\bf D53}, R1 (1996).
\bibitem{CM78} 
{ N. Cabibbo and L. Maiani}, 
{ Phys.~Lett.} {\bf B79} (1978) 109.
\bibitem{KIMM}
{ C.S. Kim and A.D. Martin},
{ Phys.~Lett.} {\bf B225} (1989) 186.
\bibitem{ISIDORI}
G. D'Ambrosio, G. Isidori and J. Portol{\'e}s, 
{ Phys.\ Lett.} {\bf 423} (1998) 385; G. Isidori and A. Retico, 
JHEP {\bf 0209} (2002) 063.
\bibitem{Adler01}
S. Adler et al., { Phys. Rev. Lett.} {\bf 79}, (1997) 2204,
{ Phys. Rev. Lett.} {\bf 84}, (2000) 3768,
[arXiv:hep-ex/0111091].
\bibitem{AI01}
G. D'Ambrosio and G. Isidori, [arXiv:hep-ph/0112135].
\bibitem{BEFAPRZU}
D. Becirevic, S. Fajfer, S. Prelovsek and J. Zupan, [arXiv:hep-ph/0211271].
\bibitem{REL}
A.J. Buras and R. Fleischer, { Phys. Rev.} {\bf D64} (2001) 115010.
A.J. Buras and R. Buras, { Phys.\ Lett.} {\bf 501} (2001) 223.
S. Bergmann and G. Perez, { Phys. Rev.} {\bf D64} (2001) 115009, 
{ JHEP} {\bf 0008} (2000) 034. 
S. Laplace, Z. Ligeti, Y. Nir and G. Perez, 
{ Phys. Rev.} {\bf D65} (2002) 094040.
\bibitem{BUCHROSL} A.J. Buras, P.H. Chankowski, J. Rosiek and {\L}.
  S{\l}awianowska, { Nucl. Phys.} {\bf B619} (2001) 434, 
{ Phys.\ Lett.} {\bf 546} (2002) 96, [arXiv:hep-ph/0210145].
\bibitem{BRMSSM}
A.J. Buras, P. Gambino, M. Gorbahn, S. J\"ager and L. Silvestrini,
{ Nucl. Phys.} {\bf B592} (2001) 55.
\bibitem{BS98}
A.J. Buras and L. Silvestrini,
{ Nucl. Phys.} {\bf B 546} (1999) 299.
\bibitem{BHI}
G. Buchalla, G. Hiller and G. Isidori, 
{ Phys. Rev.} {\bf D63} (2001) 014015.
\end{thebibliography}
\end{document}